\documentclass[a4paper]{article}
\pdfoutput=1
\usepackage{jheppub}

\usepackage[utf8]{inputenc}

\usepackage[table ]{ xcolor}

\usepackage{multirow}

\usepackage{amsmath}

\usepackage{tikz}
\usetikzlibrary{shapes,arrows}
\tikzstyle{startstop} = [rectangle,rounded corners, minimum width=3cm,minimum height=1cm,text centered, draw=black,fill=red!30]
\tikzstyle{io} = [trapezium, trapezium left angle = 70,trapezium right angle=110,minimum width=3cm,minimum height=1cm,text centered,draw=black,fill=blue!30]
\tikzstyle{process} = [rectangle,minimum width=3cm,minimum height=1cm,text centered,text width =3cm,draw=black,fill=orange!30]
\tikzstyle{decision} = [diamond,minimum width=3cm,minimum height=1cm,shape aspect=3,inner sep = 0.4pt,text centered,draw=black,fill=green!30]
\tikzstyle{arrow} = [thick,->,>=stealth]
\tikzstyle{shadow}=[preaction={fill=black,opacity=.5,transform canvas={xshift=0.5mm,yshift=-0.5mm},shading=radial,shading angle=20},fill=red]

\tikzstyle{ellipse}=[draw, rectangle, minimum width=2.8em, rounded corners=6pt,line width=0.5pt]
\tikzstyle{pxsbx}=[trapezium, trapezium left angle=75, trapezium right angle=105, minimum width=3em, text centered, draw = black, fill=white,line width=0.5pt] 
\tikzstyle{lingxing}=[draw,diamond,shape aspect=3,inner sep = 0.4pt,thick,font=\itshape,line width=0.5pt]

\usepackage{amssymb}
\usepackage{graphicx,color}

\usepackage{amsmath,amsthm}

\usepackage{mathrsfs}

\usepackage{bm}

\ifx\Ref\undefined
\newcommand{\Ref}[1]{(\ref{#1})}
\fi

\newtheorem{Theorem}{Theorem}[section]

\newcommand{\Z}{\mathbb{Z}}
\newcommand{\R}{\mathbb{R}}
\newcommand{\C}{\mathbb{C}}

\newcommand{\half}{\frac{1}{2}}

\newcommand{\ccirc}{\kern0.2ex\vcenter{\hbox{$\scriptstyle\circ$}}\kern0.2ex}

\newcommand{\Tr}{\text{Tr}}

\newcommand{\Slc}{\mathrm{SL}(2,\mathbb{C})}

\newcommand{\Su}{\mathrm{SU}(2)}

\def\be{\begin{eqnarray}}
\def\ee{\end{eqnarray}}

\newcommand{\cc}{\mathcal C}

\newcommand{\cf}{\mathcal F}
\newcommand{\cg}{\mathcal G}
\newcommand{\ch}{\mathcal H}

\newcommand{\cn}{\mathcal N}

\newcommand{\cp}{\mathcal P}

\newcommand{\cs}{\mathcal S}

  \newcommand{\Fa}{\mathfrak{A}}

  \newcommand{\Fe}{\mathfrak{E}}
\newcommand{\ff}{\mathfrak{f}}

\newcommand{\fq}{\mathfrak{q}}

\renewcommand{\a}{\alpha}
\renewcommand{\b}{\beta}
\newcommand{\g}{\gamma}
\newcommand{\G}{\Gamma}

\newcommand{\eps}{\varepsilon}

\newcommand{\sig}{\sigma}

\renewcommand{\L }{\Lambda}

\renewcommand{\t}{\tau}

\newcommand{\rmd}{\mathrm d}

\newcommand{\lt}{\left}
\newcommand{\rt}{\right}

\newcommand{\lag}{\left\langle}

\newcommand{\tr}{\mathrm{tr}}

\newcommand{\sgn}{\mathrm{sgn}}

\title{Semiclassical Limit of New Path Integral Formulation from Reduced Phase Space Loop Quantum Gravity}

\author[1,2]{Muxin Han}  

\affiliation[1]{Department of Physics, Florida Atlantic University, 777 Glades Road, Boca Raton, FL 33431-0991, USA}

\affiliation[2]{Institut f\"ur Quantengravitation, Universit\"at Erlangen-N\"urnberg, Staudtstr. 7/B2, 91058 Erlangen, Germany}

\author[2,3]{\ Hongguang Liu}

\affiliation[3]{Center for Quantum Computing, Pengcheng Laboratory, Shenzhen 518066, China}

\emailAdd{hanm(At)fau.edu}
\emailAdd{liu.hongguang(At)cpt.univ-mrs.fr}

\abstract{Recently, a new path integral formulation of Loop Quantum Gravity (LQG) has been derived in \cite{Han:2019vpw} from the reduced phase space formulation of the canonical LQG. This paper focuses on the semiclassical analysis of this path integral formulation. We show that dominant contributions of the path integral come from solutions of semiclassical equations of motion (EOMs), which reduces to Hamilton's equations of holonomies and fluxes ${h}(e),{p}^a(e)$ in the reduced phase space $\mathcal{P}_\gamma$ of the cubic lattice $\gamma$:
\be
\frac{\mathrm{d}{h}(e)}{\mathrm{d} \tau}=\{{h}(e),\, \mathbf{H}\},\quad \frac{\mathrm{d} {p}^a(e)}{\mathrm{d} \tau}=\{{p}^a(e),\, \mathbf{H}\},\nonumber
\ee 
where $\mathbf{H}$ is the discrete physical Hamiltonian. The semiclassical dynamics from the path integral becomes an initial value problem of Hamiltonian time evolution in $\mathcal{P}_\gamma$. Moreover when we take the continuum limit of the lattice $\gamma$, these Hamilton's equations reproduce correctly classical reduced phase space EOMs of gravity coupled to dust fields in the continuum, as far as initial and final states are semiclassical. Our result proves that the new path integral formulation has the correct semiclassical limit, and indicates that the reduced phase space quantization in LQG is semiclassically consistent. Based on these results, we compare this path integral formulation and the spin foam formulation, and show that this formulation has several advantages including the finiteness, the relation with canonical LQG, and being free of cosine and flatness problems.  
}


\begin{document}

\maketitle

\section{Introduction}

In recent developments of Loop Quantum Gravity (LQG), tremendous progresses have been obtained by the covariant path integral approach (see e.g. \cite{rovelli2014covariant} for summary). The covariant path integral approach of LQG focuses on transition amplitudes of LQG states (such as spin-networks). These amplitudes sum all possible evolution histories of LQG states, reflecting the idea of Feynman's path integral. Moreover the path integral approach makes it possible to bypass complications from non-polynomial Hamiltonian constraint operator, and possibly reduce difficulties in computing physical quantities in LQG. Indeed, the path integral trades the non-commutativity of quantum operators for integrals of commutative $c$-numbers, thus may reduce complicated operator manipulations to computable integrals. It is the reason why most developments of Quantum Field Theories (QFTs) are made by using path integral formulae.    

A popular path integral approach in LQG is the \emph{Spin Foam formulation} \cite{rovelli2014covariant,Perez2012}. This formulation constructs transition amplitudes of LQG on a 4-dimensional triangulations, and all these spin foam amplitudes are made by gluing elementary building blocks called vertex amplitudes, in analogy with Feynman amplitudes made by gluing vertices and propagators. This structure of spin foam amplitudes allow them to be study both analytically and numerically. Semiclassical behaviors of spin foam amplitudes, given by the large-$j$ asymptotics, have been extensively studied analytically and found close relation to Regge calculus of discrete gravity (see e.g. \cite{CFsemiclassical,semiclassical,semiclassicalEu,HZ,HZ1,HHKR,Liu:2018gfc,Han:2017xwo,Han:2018fmu,Bahr:2017eyi,Alesci:2009ys,propagator3,Han:2017isy,propagator2}). Numerical studies of spin foam amplitudes have been developed in \cite{Dona:2019dkf,Gozzini:2019kui,Dona:2018nev,Bahr:2016hwc}. Spin foams have also been related to quantum computations recently \cite{Cohen:2020jlj,2019CmPhy...2..122L,Mielczarek:2018jsh}. However extensive studies of spin foam amplitudes reveal several severe problems:
\begin{enumerate}

\item \emph{Cosine problem}: In the large-$j$ limit, the emergent (discrete) spacetime determined by spin foam amplitude with fixed semiclassical boundary state is highly non-unique in general, even when the semiclassical boundary state specifies both metric and extrinsic curvature at the boundary, while the uniqueness only happens for a single vertex amplitude \cite{Bianchi:2010mw}. Different discrete spacetimes have different 4d orientations at individual 4-simplices \cite{HZ,HZ1}. If we view the spin foam as an initial value problem, then its semiclassical time evolution from a fixed initial condition in phase space can give many different trajectories, thus is very different from classical physics. 

\item \emph{Flatness problem}: There are evidences indicating that in the large-$j$ limit, spin foam amplitudes dominate at the flat spacetime and miss all other curved spacetimes \cite{flatness,Perini:2012nd,frankflat,lowE,LowE1}. Although some other work suggests that one may modify the large-$j$ limit and/or definitions of spin foams in order to avoid the flatness problem \cite{Han:2018fmu,claudio1,Han:2017xwo}, there is still no satisfactory resolution to the problem in full generality\footnote{See also a recent numerical study toward understanding the problem \cite{Dona:2020tvv}}. 

\item \emph{Relation with canonical LQG}: The spin foam approach has been developed in parallel to the canonical approach of LQG. It is not clear how to relate spin foam amplitudes to any transition amplitude or physical inner product in the canonical LQG (see e.g. \cite{link,link1,Han:2009bb,Engle:2009ba,Alesci:2011ia,Thiemann:2013lka} for some earlier attempts). It is not clear about the unitarity of spin foam models. 

\item \emph{Divergence:} Spin foam amplitudes are divergent unless the quantum group is employed (the quantum group relates to cosmological constant \cite{QSF,HHKR}).

\item \emph{Computational complexity}: Numerical computations are currently developed only for a single vertex amplitude. Even for the vertex amplitude, the computational complexity grows very fast as the spin $j$ increases \cite{Dona:2019dkf}. The computational complexity grows exponentially when the number of 4-simplices increases. Quantum computing might help in this perspective, but it is still at a very preliminary stage.

\item \emph{Lattice dependence}: There are infinitely many spin foam amplitudes with the same boundary state. These amplitudes are defined on different triangulations (with the same boundary). It is not clear how to remove the triangulation dependence and/or how to take the continuum limit at the quantum level. Group Field Theory (GFT) provides an interesting proposal to sum over all triangulations, but it seems still difficult to extract all semiclassical smooth spacetimes from a fixed GFT partition function (while some special cases such as black holes and cosmology can indeed be extract from the general GFT formalism \cite{Oriti:2018qty,Gielen:2017eco,Oriti:2016ueo}).

\end{enumerate}

As a different approach, a new path integral formulation of LQG has been proposed recently in \cite{Han:2019vpw}. This path integral is derived from the reduced phase space formulation of canonical LQG. The reduced phase space formulation couples gravity to matter fields such as dusts or scalar fields (clock fields), followed by a deparametrization procedure, in which gravity variables are parametrized by values of clock fields, and constraints are solved classically. Results from the deparametrization are (1) the reduced phase space $\cp$ on which all phase space functions are Dirac observables free of gauge redundancy (except for the SU(2) gauge freedom when using connection variables), and (2) the dynamics is governed by a physical Hamiltonian ${\bf H}_0$ generating physical time evolution (the physical time is the value of a clock field). The reduced phase space $\cp$ of gravity-matter system can be quantized using the standard LQG technique, and result in the physical Hilbert space $\ch$. The physical Hamiltonian is promoted to a positive self-adjoint Hamiltonian operator $\hat{\bf H}$ on $\ch$. The reduced phase space quantization of LQG has been proposed conceptually in \cite{Rovelli:1990ph,Rovelli:2001bz}, and been made concrete in \cite{Dittrich:2004cb,Thiemann:2004wk,Giesel:2007wi,Giesel:2007wk,Giesel:2007wn,Giesel:2012rb} (Section \ref{BK} provides a review of the reduced phase space formulation).   

The new path integral formula in \cite{Han:2019vpw} equals to the transition amplitude of the unitary evolution generated by $\hat{\bf H}$:
\be
A_{[g],[g']}=\langle \Psi^t_{[g]}|\,\exp[-\frac{i}{\hbar}T \hat{\bf H}]\,|\Psi^t_{[g']}\rangle
\ee 
of semiclassical initial and final physical states $\Psi^t_{[g']},\Psi^t_{[g]}$. Here $\Psi^t_{[g']},\Psi^t_{[g]}$ are SU(2) gauge invariant coherent states \cite{Ashtekar:1994nx,Thiemann:2000bw} in $\ch_\g$, the physical Hilbert space on a cubic lattice $\g$. $[g],[g']$ label gauge equivalence class of initial and final data in the phase space ($g$ is the complex coordinate of the phase space). The path integral formula is derived from $A_{[g],[g']}$ by standard method: discretizing $T$ into arbitrarily large $N$ time-steps and inserting overcompletness relation of coherent states. As a result, we obtain a discrete path integral on a 4d hypercubic lattice (see Section \ref{BK} for review). 
\be
\frac{A_{[g],[g']}}{\|\Psi_{[g]}^{t}\|\,\|\Psi_{[g^{\prime}]}^{t}\|}= \int \mathrm{d} h \prod_{i=1}^{N+1} \mathrm{d} g_{i}\, \nu[g]\, e^{S[g, h] / t}\label{Agg0}
\ee
where we can extract a ``classical action'' $S[g,h]$ from the resulting path integral formula (see Section \ref{review_quantum} for details). $\int\rmd g_i \nu[g]$ integrates coherent states intermediating the quantum transition at different time steps $\t_i=\frac{i}{N}T$. $t=\ell_P^2/a^2$ is a dimensionless semiclassicality parameter, and $a$ is a length unit determining the scale at which the physics is interested. For instance, $a$ is a macroscopic unit, e.g. $a=1mm$, when we are interested in the semiclassical limit. So $\ell_P\ll a$ and $t\to0$. Eq.\Ref{Agg0} has SU(2) integrals $\int\rmd h$ since the initial and final data have SU(2) gauge freedom.  

This path integral formula is comparable to the spin foam amplitude in the coherent state representation \cite{HZ} which is frequently used for analyzing the large-$j$ behavior. On the other hand, if we choose the clock field to be a real massless scalar, Eq.\Ref{Agg0} closely relates to the spin foam model in \cite{Kisielowski:2018oiv} \footnote{Namely \Ref{Agg0} is the coherent state representation of the amplitude in \cite{Kisielowski:2018oiv}, if their derivation uses graph-preserving Hamiltonian, and $\hat{\bf H}$ is the Hamiltonian in \cite{Assanioussi:2015gka}.}. It is a matter of changing representation basis to cast the path integral \Ref{Agg0} into a shape similar to spin foams.

In this paper, we focus on the semiclassical analysis of the path integral formulation Eq.\Ref{Agg0}, i.e. the behavior as $t\to0$. By stationary phase approximation, dominant contributions of the path integral come from solutions of semiclassical equations of motion (EOMs) $\delta S=0$. These semiclassical EOMs have been derived in \cite{Han:2019vpw}, and shown to admit time continuous limit $\Delta\t=T/N\to0$, i.e. all solutions can be approximated by continuous (and differentiable) trajectories $g(\t)$ in the reduced phase space. In this paper, we show that in the time continuous limit, semiclassical EOMs derived from Eq.\Ref{Agg0} become precisely the Hamilton's equation in the reduced phase space:
\be
\frac{\rmd {h}(e)}{\rmd \t}=\{{h}(e),\, {\bf H}\},\quad \frac{\rmd {p}^a(e)}{\rmd \t}=\{{p}^a(e),\, {\bf H}\},\label{eom_intro}
\ee 
where ${h}(e),{p}^a(e)$ are holonomy and gauge covariant flux associated to the edge $e$ in $\g$. ${h}(e),{p}^a(e)$ relates to $g(e)$ by $g(e)=e^{-i p^a(e)\t^a/2}h(e)$, $\t^a=-i(\text{Pauli Matrix})^a$. $\{\ ,\ \}$ is the Poisson bracket of the reduced phase space and reduces to the holonomy-flux algebra on $\g$. ${\bf H}$ is the semiclassical limit of $\hat{\bf H}$. 

In addition, we show in Section \ref{Lattice Continuum Limit} that when we take the continuum limit of the lattice $\g$, EOMs \Ref{eom_intro} reproduce classical reduced phase space EOMs of gravity coupled to matter fields in the continuum, as far as initial and final states $\Psi_{[g']}^t,\Psi_{[g]}^t$ are semiclassical in the sense that $[g'],[g]$ is within the classically allowed regime. The classically allowed regime in the phase space satisfy certain nonholonomic constraints required by the gravity-matter system. Our result proves that the path integral formulation Eq.\Ref{Agg0} has the correct semiclassical limit, and indicates that the reduced phase space quantization in LQG is semiclassically consistent.   

Given semiclassical initial and final states and by Hamilton's equations \Ref{eom_intro}, the semiclassical dynamics from $A_{[g],[g']}$ becomes an initial value problem of Hamiltonian time evolution in the reduced phase space. Fixing the initial condition $[g']$, solution of EOMs \Ref{eom_intro}, given by the Hamiltonian flow of ${\bf H}$, is unique up to SU(2) gauge transformation. 

If semiclassical initial and final data $[g'],[g]$ are connected by the trajectory $g(\t)$ satisfying Eqs.\Ref{eom_intro}, as $t\to0$, integrals $\int\prod_{i=1}^{N+1}\rmd g_i$ in the path integral \Ref{Agg} dominate at this semiclassical trajectory:
\be
\frac{A_{[g],[g']}}{\|\Psi^t_{[g]}\|\,\|\Psi^t_{[g']}\|}=\int\rmd h\, \frac{(2\pi t)^{\cn/2}}{\sqrt{\det(-H)}}\,\nu[g(\t),h] \, e^{S[g(\t),h]/t}\,\lt[1+O(t)\rt],\label{asymp}
\ee 
where $\cn$ is the total dimension of the integral $\int\prod_{i=1}^{N+1}\rmd g_i$ in Eq.\Ref{Agg}, and $H$ is the Hessian matrix at the solution. $S[g(\t),h]$ is the action evaluated at the solution, where the continuous trajectory $g(\t)\simeq g_i$ approximates the discrete solution as $\Delta\t$ small. Here we still have $\int \rmd h$ because the initial condition $g'$ is determined by $\Psi_{[g']}^t$ up to a gauge transformation $g'\to g'^h$. If the initial and final data $[g'],[g]$ are not connected by the trajectory $g(\t)$, the amplitude is suppressed exponentially as $t\to0$.

It is interesting to make a comparison between the new path integral formulation of LQG \Ref{Agg0} to the spin foam formulation.

\begin{enumerate}

\item Our path integral formulation is free of the cosine problem. The initial state $\Psi_{[g']}^t$ determines a unique semiclassical trajectory (up to SU(e) gauge transformations) given by the Hamiltonian flow of ${\bf H}$. The asymptotic formula has a single exponential (integrated over SU(2) gauge transformations). A key reason is that here all solutions of semiclassical EOMs admit a time continuous limit. Solutions with discontinuous orientations are forbidden. 

\item Our path integral formulation is free of the flatness problem. The semiclassical EOMs \Ref{eom_intro} from the path integral reproduce the classical EOMs of the gravity-matter system, and admit all curved solutions that are physically interesting. For instance, \cite{Han:2019vpw,cospert} have demonstrated the homogeneous and isotropic cosmology and cosmological perturbation theory from solutions.

\item There is a clear link between our path integral formulation and the canonical LQG\footnote{Some advantages of relating canonical and path integral formulation can also be seen from Loop Quantum Cosmology (LQC) \cite{Ashtekar:2010ve,Henderson:2010qd}.}. The path integral \Ref{Agg} is rigorously derived from the canonical LQG. The unitarity is manifest because the path integral equals the transition amplitude of unitary evolution generated by $\hat{\bf H}$.

\item The path integral formula \Ref{Agg} is finite, because of the transition amplitude $A_{[g],[g']}$ is manifestly finite. The finiteness is irrelevant to the cosmological constant. 

\end{enumerate}

There are open issues: Computing quantum effects within the path integral formulation \Ref{Agg} relies on knowledges of the matrix elements and/or expectation values of $\hat{\bf H}$ with respect to coherent states. The non-polynomial operator $\hat{\bf H}$ may make computations highly involved. Secondly, the path integral is constructed on the lattice $\g$, it is not clear at present if we are able to remove this lattice dependence at the quantum level. So this formulation may still share issues of computational complexity and lattice dependence with the spin foam formulation, at least at the current stage. However studies of the new path integral formulation is still at very preliminary stage, and research on overcoming these issues will be carried out in the future. Some discussions are given in Section \ref{Comparison with Spin Foam Formulation}.

Many computations in this work are carried out with Mathematica on High-Performance-Computing (HPC) servers. Some intermediate steps and results contain long formulae that cannot be shown in the paper. These formulae and Mathematica codes can be downloaded from \cite{github}.

The architecture of this paper is follows: Section \ref{BK} reviews the reduced phase space formulation of LQG and the derivation of the new path integral formulation. Section \ref{Semiclassical Equations of Motion} discusses semiclassical EOMs derived from the path integral and its time continuous limit. Section \ref{Semiclassical Dynamics as Hamiltonian Evolution} shows that semiclassical EOMs are equivalent to Hamilton's equations \Ref{eom_intro}. Section \ref{Action Principle} shows that the time continuous limit of the action $S[g,h]$ gives a canonical action with the Hamiltonian ${\bf H}$, and demonstrates that the variational principle and time continuous limit are commutative when acting on $S[g,h]$. Section \ref{Lattice Continuum Limit} analyzes semiclassical EOMs in the lattice continuum limit of $\g$, and demonstrate consistency with classical gravity-matter system. Section \ref{Comparison with Spin Foam Formulation} compares the new path integral formulation with the spin foam formulation.

\section{Reduced Phase Space Formulation of LQG}\label{BK}

\subsection{Classical Framework}\label{Classical Framework}

The reduced phase space formulation couples gravity to matter fields at classical level. These matter fields are often called clock fields. In this paper, we mainly focus on two scenarios including coupling gravity to Brown-Kucha\v{r} and Gaussian dust fields \cite{Brown:1994py,Kuchar:1990vy,Giesel:2007wn,Giesel:2012rb}.

Firstly we denote by $S_{BKD}$ the action of Brown-Kucha\v{r} dust model:
\be
S_{BKD}[\rho,g_{\mu\nu},T,S^j,W_j]&=& -\frac{1}{2}\int\rmd^4x\ \sqrt{|\det(g)|}\ \rho\ [g^{\mu\nu}U_\mu U_\nu+1],\label{dustaction}\\
U_\mu&=&-\partial_\mu T+W_j\partial_\mu S^j,
\ee
where scalars $T, S^{j=1,2,3}$ form the dust coordinates of time and space to parametrize physical fields. $\rho,\ W_j$ are Lagrangian multipliers. $\rho$ is interpreted as the dust energy density. When we couple $S_{BKD}$ to gravity (or gravity coupled to some other matter fields) and carry out Hamiltonian analysis \cite{Giesel:2012rb}, we obtain following constraints:
\be
\cc^{tot}&=&\cc+\frac{1}{2}\left[\frac{P^{2} / \rho}{\sqrt{\operatorname{det}(q)}}+\sqrt{\operatorname{det}(q)} \rho\left(q^{\a \b} U_{\a} U_{\b}+1\right)\right]=0,\label{Ctot}\\
\cc^{tot}_\a&=&\cc_\a+PT_{,\a}-P_jS^j_{,\a}=0,\label{Ca}\\
\rho^2&=&\frac{P^2}{\det(q)}\lt(1+q^{\a\b}U_\a U_\b\rt)^{-1},\label{rhoP}\\
W_j&=&P_j/P,\label{WP}
\ee
where $\a,\b$ are spatial coordinate indices, $P,P_j$ are momenta conjugate to $T,S^j$, and $\cc,\cc_\a$ are Hamiltonian and diffeomorphism constraints of gravity (or gravity coupled to some other matter fields). Firstly Eq.\Ref{rhoP} can be solved by
\be
\rho=\eps\frac{P}{\sqrt{\det(q)}}\lt(1+q^{\a\b}U_\a U_\b\rt)^{-1/2}, \quad \eps=\pm1.
\ee
$\eps$ can be fixed to $\eps=1$ by physical requirement that $U$ is timelike and future pointing \cite{Giesel:2007wi}, so $\sgn(P)=\sgn(\rho)$. Inserting this solution to Eq.\Ref{Ctot} and using Eq.\Ref{WP} lead to
\be
\cc=-P\sqrt{1+q^{\a\b}\cc_\a \cc_\b/P^2}.
\ee
Thus $-\sgn(\cc)=\sgn(P)=\sgn(\rho)$. When we consider dust coupling to pure gravity, we must have $\cc<0$ and the physical dust $\rho,P>0$ to fulfill the energy condition as in \cite{Brown:1994py}. However, we may couple some additional matter fields (e.g. scalars, fermions, gauge fields etc) to make $\cc>0$, then $\rho,P<0$ correspond to the phantom dust as in \cite{Giesel:2007wn,Giesel:2007wi}. The case of phantom dust may not violate the usual energy condition due to the presence of additional matter fields. We can solve $P,P_j$ from Eqs.\Ref{Ctot} and \Ref{Ca}
\be
&&P=\begin{cases} h &\  \text{physical dust}, \\
-h &\ \text{phantom dust},
\end{cases} \quad h=\sqrt{\cc^2-q^{\a\b}\cc_\a \cc_\b},\label{P=-h}\\
&&P_j=-S^\a_j\lt(\cc_\a-hT_{,\a}\rt)
\ee
which are strongly Poisson commutative constraints. $S^\a_j$ is the inverse matrix of $\partial_\a S^j$ ($\a=1,2,3$). In deriving above constraints, we find at an intermediate step that $P^2 = \cc^2- q^{\a\b}\cc_\a \cc_\b>0$ constraints the argument of the square root to be positive. Moreover the physical dust requires $\cc<0$ while the phantom dust requires $\cc>0$.

We use $A^a_\a(x),E^\a_a(x)$ to be canonical variables of gravity, where $A^a_\a(x)$ is the Ashtekar-Barbero connection and $E^\a_a(x)=\sqrt{\det q}\, e^\a_a(x)$ is the densitized triad. $a=1,2,3$ is the Lie algebra index of su(2). Gauge invariant Dirac observables are constructed relationally by parametrizing $(A,E)$ with values of dust fields $T(x)\equiv\t,S^j(x)\equiv\sig^j$, i.e. $A_j^a(\sig,\t)=A_j^a(x)|_{T(x)\equiv\t,\,S^j(x)\equiv\sig^j}$ and $E^j_a(\sig,\t)=E^j_a(x)|_{T(x)\equiv\t,\,S^j(x)\equiv\sig^j}$, where $\sig,\t$ are physical space and time coordinates in the dust reference frame. Here $j=1,2,3$ is the dust coordinate index (e.g. $A_j=A_\a S^\a_j$).

Both $A_j^a(\sig,\t)$ and $E^j_a(\sig,\t)$ are free of diffeomorphism and Hamiltonian constraints. They satisfy the standard Poisson bracket in the dust frame: 
\be
\{E^i_a(\sig,\t),A_j^b(\sig',\t)\}=\frac{1}{2}\kappa \b\ \delta^{i}_j\delta^b_a\delta^{3}(\sig,\sig')
\ee 
where $\b$ is the Barbero-Immirzi parameter and $\kappa=16\pi G$. The reduced phase space $\cp$ of $A_j^a(\sig,\t),E^j_a(\sig,\t)$ is free of Hamiltonian and diffeomorphism constraints. All SU(2) gauge invariant phase space functions are Dirac observables.

The evolution in physical time $\t$ is generated by the classical physical Hamiltonian ${\bf H}_0$ given by integrating $h$ on the constant $T=\t$ slice $\cs$. The constant $\t$ slice $\cs$ is coordinated by the value of dust scalars $S^j=\sig^j$ thus is referred to as the dust space \cite{Giesel:2007wn,Giesel:2012rb}. From Eq.\Ref{P=-h}, we find that ${\bf H}_0$ is negative for physical dust while is positive for phantom dust. We flip the direction of the time flow $\t\to -\t$ thus ${\bf H}_0 \to -{\bf H}_0$ for physical dust so we have a positive Hamiltonians in every case:
\be
{\bf H}_0=\int_\cs\rmd^3\sig\, \sqrt{\cc(\sig,\t)^2-\frac{1}{4}\sum_{a=1}^3\cc_a(\sig,\t)^2}.
\ee 
Here $\cc$ and $\cc_a=2e_a^\a \cc_\a$ are parametrized in the dust frame. In terms of $A_j^a(\sig,\t)$ and $E^j_a(\sig,\t)$:
\be
\cc&=&\frac{1}{\kappa}\lt[F^a_{jk}-\lt({\b^2+1}\rt)\eps_{ade}K^d_{j}K^e_{k}\rt]\eps_{abc}\frac{E^j_bE^k_c}{\sqrt{\det(q)}}+\frac{2\L}{\kappa}\sqrt{\det(q)}\\
\cc_a&=&\frac{4}{\kappa\b}F^b_{jk} \frac{E^j_aE^k_b}{\sqrt{\det(q)}}.
\ee  
$\L$ is the cosmological constant.

Coupling gravity to Gaussian dust model can be analyzed similarly, so we don't present the details here (while details can be found in \cite{Giesel:2012rb}). As a result the physical Hamiltonian has a simpler expression 
\be
\mathbf{H}_0=\int_\cs\rmd^3\sig\, \cc(\sig,\t). \label{gaussd}
\ee
In order to put discussions of both the Brown-Kucha\v{r} and Gaussian dusts in a unified manner, we express the physical Hamiltonian as the following:
\be
\mathbf{H}_0&=&\int_\cs\rmd^3\sig\, h(\sig,\t),\label{ham1}\\
 h(\sig,\t)&=&\sqrt{\cc(\sig,\t)^2-\frac{\a}{4}\sum_{a=1}^3\cc_a(\sig,\t)^2},\quad \begin{cases} 
\a =1& \text{Brown-Kucha\v{r} dust},\\
\a =0& \text{Gaussian dust}.
\end{cases} \nonumber
\ee 
The physical Hamiltonian ${\bf H}_0$ is manifestly positive in Eq.\Ref{ham1}. When $\cc<0$, Eq.\Ref{ham1} is different from Eq.\Ref{gaussd} by an overall minus sign, thus reverses the time flow $\t\to-\t$ for the Gaussian dust.

In both scenarios, the physical Hamiltonian $\mathbf{H}_0$ generates the $\t$-time evolution:
\be
\frac{\rmd f}{\rmd\t}=\lt\{ f, \mathbf{H}_0\rt\},
\ee
for all phase space function $f$ of $A_j^a(\sig,\t)$ and $E^j_a(\sig,\t)$. In particular, the Hamilton's equations are
\be
\frac{\rmd A^a_j(\sig,\t)}{\rmd\t}=-\frac{\kappa\b}{2}\frac{\delta\mathbf{H}_0}{\delta E^j_a(\sig,\t)},\quad \frac{\rmd E^j_a(\sig,\t)}{\rmd\t}=\frac{\kappa\b}{2}\frac{\delta\mathbf{H}_0}{\delta A^a_j(\sig,\t)}.\label{hamitoncon0}
\ee

Functional derivatives on the right-hand sides of Eq.\Ref{hamitoncon0} can be computed by
\be
\delta{\bf H}_0=\int_\cs\rmd^3\sig \lt(\frac{\cc}{h}\delta \cc -\frac{\a}{4}\frac{\cc_a}{h}\delta\cc_a\rt),
\ee
where ${C}/{h}$ is negative (positive) for physical (phantom) dust. Compare $\delta{\bf H}$ to the variation of Hamiltonian $H_{GR}$ of pure gravity in absence of dust motivates us to view the following as physical lapse function and shift vector 
\be
N=\frac{\cc}{h}, \quad N_a=-\frac{\a}{4}\frac{\cc_a}{h}.
\ee
Therefore $N$ is negative (positive) for the physical (phantom) dust. Negative $N$ for the physical dust relates to the flip $\t\to-\t$ for making Hamiltonian positive. 

In the gravity-dust models, we resolve the Hamiltonian and diffeomorphism constraints classically, while the SU(2) Gauss constraint $\cg_a(\sig,\t)=D_j E^j_a(\sig,\t)=0$ still has to be imposed to the phase space. In addition, non-holonomic constraints are imposed to the phase space: $\cc(\sig,\t)^2-\frac{\a}{4}\sum_{a=1}^3\cc_a(\sig,\t)^2\geq 0$ and $\cc<0$ for physical dust ($\cc>0$ for phantom dust).

These constraints are preserved by the time evolution for gravity coupled to the Brown-Kucha\v{r} dust. Indeed, firstly the time evolution cannot break Gauss constraint since $\lt\{\cg_a(\sig,\t),\,{\bf H}_0\rt\}=0$. Secondly both $h(\sig,\t)$ and $\cc_j(\sig,\t)=\half e^a_j\cc_a(\sig,\t)$ are conserved densities on the Gauss constraint surface \cite{Giesel:2007wn}:
\be
\frac{\rmd h(\sig,\t)}{\rmd \t}=\lt\{h(\sig,\t),\,{\bf H}_0\rt\}=0,\quad \frac{\rmd \cc_j(\sig,\t)}{\rmd \t}=\lt\{\cc_j(\sig,\t),\,{\bf H}_0\rt\}=0
\ee
Therefore $\cc(\sig,\t)^2-\frac{1}{4}\sum_{a=1}^3\cc_a(\sig,\t)^2\geq 0$ is conserved in the time evolution. About the other non-holonomic constraint $\cc<0$ ($\cc>0$), suppose $\cc<0$ ($\cc>0$) was violated in the time evolution, there would exist a certain time $\t_0$ that $\cc(\sig,\t_0)=0$, but then $\cc(\sig,\t)^2-\frac{1}{4}\sum_{a=1}^3\cc_a(\sig,\t)^2$ would becomes negative if $\cc_j(\sig,\t)\neq 0$, contradicting the conservation of $h(\sig,\t)$ and the other nonholonomic constraint. If the conserved $\cc_j(\sig,\t)=0$, $h(\sig,\t)^2=\cc(\sig,\t)^2$ is conserved so cannot evolve from nonzero to zero. For gravity coupled to the Gaussian dust, $\cc_j(\sig,\t)$ is conserved. $h(\sig,\t)$ and $\cc(\sig,\t)$ are conserved only when $\cc_j(\sig,\t)=0$. $\cc<0$ ($\cc>0$) may be violated in the time evolution for gravity coupled to the Gaussian dust if $\cc_j(\sig,\t)\neq 0$.

In our following discussion, we focus on pure gravity coupling to dusts, thus we only work with physical dusts in order not to violating the energy condition.


\subsection{Quantization, Transition Amplitude, and Coherent State Path Integral}\label{review_quantum}

We construct a fixed cubic lattice $\g$ which partitions the dust space $\cs$. In this work, we consider $\cs$ is compact and has no boundary so that $\g$ is a finite lattice. We denote by $E(\g)$ and $V(\g)$ sets of (oriented) edges and vertices in $\g$. By the dust coordinate on $\cs$, we assign every edge a constant coordinate length $\mu$. $\mu\to 0$ relates to the lattice continuum limit. Every vertex $v\in V(\g)$ is 6-valent. At $v$ there are 3 outgoing edges $e_I(v)$ ($I=1,2,3$) and 3 incoming edges $e_I(v-\mu \hat{I})$ where $\hat{I}$ is the coordinate basis vector along the $I$-th direction. It is sometimes convenient to orient all 6 edges at $v$ to be outgoing from $v$, and denote 6 edges by $e_{v;I,s}$ ($s=\pm$):
\be
e_{v;I,+}=e_I(v),\quad e_{v;I,-}=e_I(v-\mu \hat{I})^{-1}.
\ee

We regularize canonical variables $A^a_j(\sig,\t),E^j_a(\sig,\t)$ on the lattice $\g$, by defining holonomy $h(e)$ and gauge covariant flux $p^a(e)$ at every $e\in E(\g)$:
\be
h(e)&:=&\cp \exp \int_{e}A,\nonumber\\
p^a(e)&:=&-\frac{1}{2\b a^2}\tr\lt[\t^a\int_{S_e}\eps_{ijk}\rmd \sig^i\wedge\rmd \sig^j\ h\lt(\rho_e(\sig)\rt)\, E_b^k(\sig)\t^b\, h\lt(\rho_e(\sig)\rt)^{-1}\rt],\label{hpvari}
\ee 
where $A=A^a\t^a/2$ and $\t^a=-i(\text{Pauli matrix})^a$. $S_e$ is a 2-face intersecting $e$ in the dual lattice $\g^*$. $\rho_e$ is a path starting at the source of $e$ and traveling along $e$ until $e\cap S_e$, then running in $S_e$ until $\vec{\sig}$. $a$ is a length unit for making $p^a(e)$ dimensionless. Note that because $p^a(e)$ is gauge covariant flux, we have
\be
p^{a}\left(e_{v ; I,-}\right)=\frac{1}{2} \operatorname{Tr}\left[\tau^{a} h\left(e_{v-\hat{I} ; I,+}\right)^{-1} p^{b}\left(e_{v-\hat{I} ; I,+}\right) \tau^{b} h\left(e_{v-\hat{I} ; I,+}\right)\right].
\ee
The Poisson algebra of $h(e)$ and $p^a(e)$ are called the holonomy-flux algebra:
\be
\left\{h(e), h\left(e^{\prime}\right)\right\} &=&0 ,\label{handh}\\
\left\{p^{a}(e), h\left(e^{\prime}\right)\right\} &=&\frac{\kappa}{a^{2}} \delta_{e, e^{\prime}} \frac{\tau^{a}}{2} h\left(e^{\prime}\right) ,\label{pandtheta}\\
\left\{p^{a}(e), p^{b}\left(e^{\prime}\right)\right\} &=&-\frac{\kappa}{a^{2}} \delta_{e, e^{\prime}} \varepsilon_{a b c} p^{c}\left(e^{\prime}\right),\label{pandp}
\ee
$h(e)$ and $p^a(e)$ parametrize the reduced phase space $\cp_\g$ for the theory discretized on $\g$.

The LQG quantization defines the Hilbert space $\ch_\g$ spanned by gauge invariant (complex valued) functions of all $h(e)$'s on $\g$, and is a proper subspace of $\ch_\g^0=\otimes_e L^2(\Su)$. $\ch_\g$ is the physical Hilbert space free of constraint because it quantizes the reduced phase space. $\hat{h}(e)$ becomes multiplication operators on functions in $\ch_\g^0$. $\hat{p}^a(e)=i t\,\hat{R}_e^a/2$ where $\hat{R}_e^a$ is the right invariant vector field on SU(2): $R^a f(h)=\frac{\rmd}{\rmd \eps}\big|_{\eps=0} f(e^{\eps\t^a}h)$. $t=\ell^2_p/a^2$ is a dimensionless semiclassicality parameter ($\ell^2_p=\hbar\kappa$). $\hat{h}(e),\hat{p}^a(e)$ satisfy the commutation relations:
\be
\lt[\hat{h}(e),\hat{h}(e')\rt] &=&0\nonumber\\
\lt[\hat{p}^a(e),\hat{h}(e')\rt] &=&i t \delta_{e,e'} \frac{\t^a}{2} {h}(e')\nonumber\\
\lt[\hat{p}^a(e),\hat{p}^b(e')\rt]&=&-it \delta_{e,e'} \eps_{abc} {p}^c(e'), \label{ph}
\ee
as quantization of the holonomy-flux algebra.

The (non-graph-changing) physical Hamiltonian operators $\hat{\bf H}$ are given by \cite{Giesel:2007wn}:
\be
\hat{\mathbf{H}}&=&\sum_{v\in V(\g)}\hat{H}_v,\quad \hat{H}_v:=\lt[\hat{M}_-^\dagger(v) \hat{M}_-(v)\rt]^{1/4},\label{physHam}\\ 
\hat{M}_-(v)&=&\hat{C}_{v}^{\ \dagger}\hat{C}_{v}-\frac{\a}{4}\sum_{a=1}^3\hat{C}_{a,v}^{\ \dagger}\hat{C}_{a,v},\quad \a=\begin{cases}
1,&\text{Brown-Kucha\v{r} dust,}\\
0,&\text{Gaussian dust.}
\end{cases}
\ee 
In our notation, ${\bf H}_0=\int_\cs\rmd^3\sig\, h$, $\cc$, and $\cc_{a}$ are the Hamiltonian, scalar constraint, and vector constraint in the continuum. ${\bf H}=\sum_v H_v$, $C_v$, and $C_{a,v}$ are their discretizations on $\g$, while $\hat{\bf H}=\sum_v \hat{H}_v$, $\hat{C}_v$, and $\hat{C}_{a,v}$ are quantizations of ${\bf H}$, $C_v$, and $C_{a,v}$:
\be
\hat{C}_{0,v}&=&\frac{2}{i\b\kappa\ell_p^2}\sum_{s_1,s_2,s_3=\pm1}s_1s_2s_3\ \eps^{I_1I_2I_3}\ \mathrm{Tr}\Bigg(\hat{h}(\a_{v;I_1s_1,I_2s_2}) \hat{h}(e_{v;I_3s_3})\Big[\hat{h}(e_{v;I_3s_3})^{-1},\hat{V}_v\Big] \Bigg)\label{C}\\
\hat{C}_{a,v}&=&\frac{8}{i\b^2\kappa\ell_p^2}\sum_{s_1,s_2,s_3=\pm1}s_1s_2s_3\ \eps^{I_1I_2I_3}\ \mathrm{Tr}\Bigg(\t^a \hat{h}(\a_{v;I_1s_1,I_2s_2}) \hat{h}(e_{v;I_3s_3})\Big[\hat{h}(e_{v;I_3s_3})^{-1},\hat{V}_v\Big] \Bigg)\\
\hat{C}_v&=&\hat{C}_{0,v}+\frac{1+\b^2}{2}\hat{C}_{L,v}+\frac{2\L}{\kappa}\hat{ V}_v,\quad\quad \hat{K}=\frac{i}{\hbar\b^2}\lt[\sum_{v\in V(\g)}\hat{C}_{0,v},\sum_{v\in V(\g)}V_v\rt]\nonumber\\
\hat{C}_{L,v}&=&-\frac{16}{\kappa\lt(i\b\ell_p^2\rt)^3}\sum_{s_1,s_2,s_3=\pm1}s_1s_2s_3\ \eps^{I_1I_2I_3}\label{HCO}\\
&&\mathrm{Tr}\Bigg( \hat{h}(e_{v;I_1s_1})\Big[\hat{h}(e_{v;I_1s_1})^{-1},\hat{K}\Big]\ \hat{h}(e_{v;I_2s_2})\Big[\hat{h}(e_{v;I_2s_2})^{-1},\hat{K}\Big]\ \hat{h}(e_{v;I_3s_3})\Big[\hat{h}(e_{v;I_3s_3})^{-1},\hat{V}_v\Big]\ \Bigg).\nonumber
\ee
where $\L$ is the cosmological constant and $\hat{V}_v$ is the volume operator at $v$:
\be
\hat{V}_v&=&\lt(\hat{Q}_v^2\rt)^{1/4},\\ 
\hat{Q}_v
&=&-i\lt(\frac{\b\ell_P^2}{4}\rt)^3\eps_{abc}\frac{R^a_{e_{v;1+}}-R^a_{e_{v;1-}}}{2}\frac{R^b_{e_{v;2+}}-R^b_{e_{v;2-}}}{2}\frac{R^c_{e_{v;3+}}-R^c_{e_{v;3-}}}{2}\nonumber\\
&=&\b^3a^6\eps_{abc}\frac{\hat{p}^a({e_{v;1+}})-\hat{p}^a({e_{v;1-}})}{4}\frac{\hat{p}^b({e_{v;2+}})-\hat{p}^b({e_{v;2-}})}{4}\frac{\hat{p}^c({e_{v;3+}})-\hat{p}^c({e_{v;3-}})}{4}\label{Qv}
\ee 

The Hamiltonian operator $\hat{\mathbf{H}}$ is positive semi-definite and self-adjoint because $\hat{M}_-^\dagger(v) \hat{M}_-(v)$ is manifestly positive semi-definite and Hermitian, therefore admits a self-adjoint extension (Friedrich extension). 

Classical discrete $C_v$, and $C_{a,v}$ are obtained from Eqs.\Ref{C} - \Ref{HCO} by mapping operators to their classical counterparts and $[\hat{f}_1,\hat{f}_2]\to i\hbar\{f_1,f_2\} $. Hence classical discrete physical Hamiltonian ${\bf H}$ is given by
\be
{\bf H}=\sum_{v\in V(\g)} H_v,\quad H_v=\sqrt{\lt|C_v^2-\frac{\a}{4}\sum_{a=1}^3 C_{a,v}^2\rt|}.\label{physHamcl}
\ee
The absolute value in the square-root results from that ${\bf H}$ is the classical limit of $\hat{\bf H}$ defined on the entire $\ch_\g$ disregarding nonholonomic constraints in particular $\cc^2-\frac{\a}{4}\sum_{a=1}^3\cc_a^2\geq 0$ for $\a=1$.

An interesting quantity for quantum dynamics is the transition amplitude
\be
A_{[g],[g']}=\langle \Psi^t_{[g]}|\,\exp\lt[-\frac{i}{\hbar}T \hat{\bf H}\rt]\,|\Psi^t_{[g']}\rangle
\ee
For the purpose of semiclassical analysis, we focus on the semiclassical initial and final states $\Psi^t_{[g']}, \Psi^t_{[g]}$ which are gauge invariant coherent states defined in \cite{Thiemann:2000bw,Thiemann:2000ca}:
\be
\Psi^t_{[g]}(h)
&=&\int_{\mathrm{SU(2)}^{|V(\g)|}}\rmd h\prod_{e\in E(\g)}{\psi}^t_{h_{s(e)}^{-1}g(e)h_{t(e)}}\lt(h(e)\rt),\quad \rmd h=\prod_{v\in V(\g)}\rmd\mu_H(h_v).\label{gaugeinv}
\ee
where $\rmd\mu_H(h_v)$ is the Haar measure on SU(2). The gauge invariant coherent state is labelled by gauge equivalence class $[g]$ generated by $g(e)\sim g^h(e)= h_{s(e)}^{-1}g(e)h_{t(e)}$ at all $e$. Here $g(e)$ is an $\Slc$ group element. $\psi^{t}_{g(e)}\lt(h(e)\rt)$ is the complexifier coherent state on the edge $e$:
\be
\psi^{t}_{g(e)}\lt(h(e)\rt)
&=&\sum_{j_e\in\Z_+/2\cup\{0\}}(2j_e+1)\ e^{-tj_e(j_e+1)/2}\chi_{j_e}\lt(g(e)h(e)^{-1}\rt),\label{coherent}
\ee
where $g(e)$ is complex coordinate of $\cp_\g$ and relates to $h(e),p^a(e)$ by\footnote{For any polynomial $\mathrm{Pol}[\hat{h}(e),\hat{p}^a(e)]$ of $\hat{h}(e),\hat{p}^a(e)$, the coherent state expectation value is semiclassical: $\langle \psi_{g(e)}^t |\mathrm{Pol}[\hat{h}(e),\hat{p}^a(e)]|\psi_{g(e)}^t\rangle=\mathrm{Pol}[{h}(e),{p}^a(e)]+O(t)$ where ${h}(e),{p}^a(e)$ on the right hand side relate to $g(e)$ by Eq.\Ref{gthetap} \cite{Thiemann:2000bx}.}
\be
g(e)=e^{-ip_a(e)\t_a/2}h(e)=
e^{-ip^a(e)\t^a/2}e^{\theta^a(e)\t^a/2}, \quad p^a(e),\ \theta^a(e)\in\R^3.\label{gthetap}
\ee

Applying Eq.\Ref{gaugeinv} and using a discretization of time $T=N\Delta\t$ with large $N$ and infinitesimal $\Delta\t$, 
\be
A_{[g],[g']}&=&\int\rmd h\lag\psi^t_{g}\rt|\lt[e^{ -\frac{i}{\hbar}\Delta\t \hat{\mathbf{H}}}\rt]^N |{\psi}^t_{{g'}^{h}}\rangle,\\
&=&\int\rmd h\prod_{i=1}^{N+1}\mathrm{d}g_{i}\,\langle\psi^t_{g}|\tilde{\psi}^t_{g_{N+1}}\rangle\langle \tilde{\psi}^t_{g_{N+1}}\big|e^{ -\frac{i\Delta\t}{\hbar} \hat{\mathbf{H}}}\big|\tilde{\psi}^t_{g_{N}}\rangle
\langle \tilde{\psi}^t_{g_{N}}\big|e^{ -\frac{i\Delta\t}{\hbar}\hat{\mathbf{H}}}\big|\tilde{\psi}^t_{g_{N-1}}\rangle\cdots\nonumber\\
&&\quad \cdots\ 
\langle \tilde{\psi}^t_{g_2}\big|e^{ -\frac{i\Delta\t}{\hbar}\hat{\mathbf{H}}}\big|\tilde{\psi}^t_{g_1}\rangle\langle\tilde{\psi}^t_{g_1}|{\psi}^t_{g'{}^{h}}\rangle \label{smallsteps}
\end{eqnarray}
where we have inserted $N+1$ overcompleteness relations of normalized coherent state $\tilde{\psi}^t_{g}=\otimes_e{\psi}^t_{g(e)}/||{\psi}^t_{g(e)}||$:
\be
\int\rmd g_i\ |\tilde{\psi}^{t}_{g_i}\rangle\langle\tilde{\psi}^{t}_{g_i}|=1_{\ch_\g^0},\quad \rmd g_i=\lt(\frac{c}{t^3}\rt)^{|E(\g)|}\prod_{e\in E(\g)}\rmd\mu_H(h_i(e))\,\rmd^3p_i(e),\quad i=1,\cdots,N-1.
\ee

A path integral formula is derived in \cite{Han:2019vpw} from the above expression of $A_{[g],[g']}$:
\be
A_{[g],[g']}=\left\|\psi_{g}^{t}\right\|\left\|\psi_{g^{\prime}}^{t}\right\| \int \mathrm{d} h \prod_{i=1}^{N+1} \mathrm{d} g_{i}\, \nu[g]\, e^{S[g, h] / t}\label{Agg}
\ee 
where the ``effective action'' $S[g,h]$ is given by
\be
S[g, h]&=&\sum_{i=0}^{N+1} K\left(g_{i+1}, g_{i}\right)-\frac{i \kappa}{a^{2}} \sum_{i=1}^{N} \Delta \tau\left[\frac{\langle\psi_{g_{i+1}}^{t}|\hat{\mathbf{H}}| \psi_{g_{i}}^{t}\rangle}{\langle\psi_{g_{i+1}}^{i} | \psi_{g_{i}}^{t}\rangle}+i \tilde{\varepsilon}_{i+1, i}\left(\frac{\Delta \tau}{\hbar}\right)\right],\label{Sgh}\\
K\left(g_{i+1}, g_{i}\right)&=&\sum_{e \in E(\gamma)}\left[z_{i+1, i}(e)^{2}-\frac{1}{2} p_{i+1}(e)^{2}-\frac{1}{2} p_{i}(e)^{2}\right]
\ee
with $g_{0} \equiv g^{\prime h},\ g_{N+2} \equiv g$, and $\nu[g]$ is a measure factor. $\tilde{\varepsilon}_{i+1, i}\left(\frac{\Delta \tau}{\hbar}\right)\to 0$ as $\Delta\t\to0$ and is negligible. In the above, $z_{i+1,i}(e)$ and $x_{i+1,i}(e)$ are given by 
\be
z_{i+1,i}(e)&=& \mathrm{arccosh}\lt(x_{i+1,i}(e)\rt),\quad x_{i+1,i}(e)=\half\tr\lt[g_{i+1}(e)^\dagger g_{i}(e)\rt].
\ee

The path integral Eq.\Ref{Agg} is constructed with discrete time and space, and is a well-define integration formula for the transition amplitude $A_{[g],[g']}$ as long as $\Delta\t$ is arbitrarily small but finite. The time translation of $\g$ with finite $\Delta\t$ makes a hypercubic lattice in 4 dimensions, on which the path integral is defined. There is no issue of any divergence in this path integral formulation of LQG, since it is derived from a well-defined transition amplitude. 

\section{Semiclassical Equations of Motion}\label{Semiclassical Equations of Motion}

\subsection{Discrete Equations of Motion}

The main part of this work is to study the semiclassical limit $t\to0$ (or $\ell_P\ll a$) of the transition amplitude $A_{[g],[g']}$. By Eq.\Ref{Agg} and the stationary phase approximation, dominant contributions to $A_{[g],[g']}$ as $t\to0$ come from semiclassical trajectories satisfying the semiclassical equations of motion (EOMs). 

Semiclassical EOMs has been derived in \cite{Han:2019vpw} by the variational principle $\delta S[g,h]=0$ and expressed in the following form:

\begin{itemize}

\item For $i=1,\cdots,N$, at every edge $e\in E(\g)$,
\be
\frac{1}{\Delta\t}\lt[\frac{z_{i+1,i}(e)\,\tr\lt[\t^a g_{i+1}(e)^\dagger g_i(e)\rt]}{\sqrt{x_{i+1,i}(e)-1}\sqrt{x_{i+1,i}(e)+1}}-\frac{p_i(e)\,\tr\lt[\t^a g_{i}(e)^\dagger g_i(e)\rt]}{\sinh(p_i(e))}\rt]\nonumber\\
=\frac{i\kappa }{a^2}\frac{\partial}{\partial \varepsilon_{i}^{a}(e)} \frac{\langle\psi_{g_{i+1}^{\varepsilon}}^{t}|\hat{\mathbf{H}}| \psi_{g_{i}^{\varepsilon}}^{t}\rangle}{\langle\psi_{g_{i+1}^{\varepsilon}}^{t} | \psi_{g_{i}^{\varepsilon}}^{t}\rangle}\Bigg|_{\vec{\eps}=0}\label{eoms1}
\ee
where $g^\eps(e)=g(e) e^{\eps^a(e)\t^a}$ ($\eps^a(e)\in\C$) is a holomorphic deformation.

\item For $i=2,\cdots,N+1$, at every edge $e\in E(\g)$,
\be
\frac{1}{\Delta\t}\lt[\frac{z_{i,i-1}(e)\,\tr\lt[\t^a g_{i}(e)^\dagger g_{i-1}(e)\rt]}{\sqrt{x_{i,i-1}(e)-1}\sqrt{x_{i,i-1}(e)+1}}-\frac{p_i(e)\,\tr\lt[\t^a g_{i}(e)^\dagger g_i(e)\rt]}{\sinh(p_i(e))}\rt]\nonumber\\
=-\frac{i\kappa }{a^2}\frac{\partial}{\partial \bar{\varepsilon}_{i}^{a}(e)} \frac{\langle\psi_{g_{i}^{\varepsilon}}^{t}|\hat{\mathbf{H}}| \psi_{g_{i-1}^{\varepsilon}}^{t}\rangle}{\langle\psi_{g_{i}^{\varepsilon}}^{t} | \psi_{g_{i-1}^{\varepsilon}}^{t}\rangle}\Bigg|_{\vec{\eps}=0}.\label{eoms2}
\ee

\item The closure condition at every vertex $v\in V(\g)$ for initial data:
\be
-\sum_{e, s(e)=v}p_1^a(e)+\sum_{e, t(e)=v}\L^a_{\ b}\lt(\vec{\theta}_1(e)\rt)\,p_1^b(e)=0.\label{closure0}
\ee
where $\L^a_{\ b}(\vec{\theta})\in\mathrm{SO}(3)$ is given by $e^{\theta^c\t^c/2}\t^a e^{-\theta^c\t^c/2}=\L^a_{\ b}(\vec{\theta})\t^b$.

\end{itemize}
\noindent
The initial and final conditions are given by $g_{1}=g'^h$ and $g_{N+1}=g$. Here the gauge transformation $h$ is arbitrary. Eqs.\Ref{eoms1} and \Ref{eoms2} come from $\delta S/\delta g=0$ and $\delta S/\delta\bar{g}=0$, while Eq.\Ref{closure0} comes from $\delta S/\delta h=0$. These semiclassical EOMs govern the semiclassical dynamics of LQG in the reduced phase space formulation.

Semiclassical EOMs \Ref{eoms1} - \Ref{closure0} are derived with finite $\Delta\t$. We prefer to derive EOMs from the path integral Eq.\Ref{Agg} with discrete time and space, because Eq.\Ref{Agg} is a well-define integration formula for the transition amplitude. 

The small-step transitions $\langle \tilde{\psi}^t_{g_{i+1}}|\exp \lt( -\frac{i}{\hbar}\Delta\t \hat{\mathbf{H}}\rt)|\tilde{\psi}^t_{g_{i}}\rangle$ in Eq.\Ref{smallsteps} are dominated by overlaps $\langle \tilde{\psi}^t_{g_{i+1}}|\tilde{\psi}^t_{g_{i}}\rangle$ as $\Delta\t$ is arbitrarily small. $|\langle \tilde{\psi}^t_{g_{i+1}}|\tilde{\psi}^t_{g_{i}}\rangle|$ decays exponentially fast to zero unless $g_{i+1}$ is within a small neighborhood at $g_i$ of radius $\sqrt{t}$ \cite{Thiemann:2000ca} (a summary can be found in \cite{Giesel:2006um}). Therefore for sufficiently large $N$, the dominant contribution to $A_{[g],[g']}$ in Eq.\Ref{Agg} comes from integral over the neighborhood where all $g_{i+1}$ are close to $g_i$ with distance of $O(\sqrt{t})$. This neighborhood becomes arbitrarily small as $t\to0$. Within this neighborhood, both quantities in square brackets in Eqs.\Ref{eoms1} and \Ref{eoms2} have a single isolated zero at $g_i=g_{i+1}$ (Lemma 4.1 in \cite{Han:2019vpw}). Therefore $\Delta\t\to0$ forces $g_i\to g_{i+1}$, given that right-hand sides of Eqs.\Ref{eoms1} and \Ref{eoms2} are always finite \cite{Han:2019vpw}. So any solution of Eqs.\Ref{eoms1} and \Ref{eoms2} can be approximated arbitrarily well by the continuous function $g_i\simeq g(\t)$, as $\Delta\t$ arbitrarily small. In the following we apply this approximation, replace all $g_i$ by continuous function $g(\t)$, and take the time continuous limit $\Delta\t\to0$ of Eqs.\Ref{eoms1} and \Ref{eoms2}.

\subsection{Time Continuous Limit}

The time continuous limit leads to $g_{i+1}\to g_{i}=g(\t)$, so that matrix elements $\langle\psi_{g_{i}^{\varepsilon}}^{t}|\hat{\mathbf{H}}| \psi_{g_{i-1}^{\varepsilon}}^{t}\rangle$ on right-hand sides of Eqs.\Ref{eoms1} - \Ref{eoms2} reduces to the expectation values $\langle\psi_{g^{\varepsilon}}^{t}|\hat{\mathbf{H}}| \psi_{g^{\varepsilon}}^{t}\rangle$ as $\Delta\t\to0$ (see \cite{Han:2019vpw} for proving that $g_{i+1}\to g_{i}$ commutes with holomorphic derivatives). Coherent state expectation values of $\hat{\bf H}$ have correct semiclassical limit\footnote{Firstly we can apply the semiclassical perturbation theory of \cite{Giesel:2006um} to $\hat{O}\equiv\hat{{H}}_v^4$ (recall Eq.\Ref{physHam}) and all $\hat{O}^n$ ($n>1$): $\langle\tilde{\psi}_{g}^{t}|\hat{O}^n| \tilde{\psi}_{g}^{t}\rangle={O}[g]^n+O(t)$. Then by Theorem 3.6 of \cite{Thiemann:2000bx}, $\lim_{t\to0}\langle\tilde{\psi}_{g}^{t}|f(\hat{O})| \tilde{\psi}_{g}^{t}\rangle=f({O}[g])$ for any any Borel measurable function on $\R$ such that $\langle\tilde{\psi}_{g}^{t}|f(\hat{O})^\dagger f(\hat{O})| \tilde{\psi}_{g}^{t}\rangle<\infty$.}
\be
\lim_{t\to0}\langle\tilde{\psi}_{g}^{t}|\hat{\mathbf{H}}| \tilde{\psi}_{g}^{t}\rangle={\bf H}[g]
\ee
where ${\bf H}[g]$ is the classical discrete Hamiltonian \Ref{physHamcl} evaluated at $p^a(e),h(e)$ determined by $g(e)$ in Eq.\Ref{gthetap}. Note that deriving semiclassical behavior of $\langle\tilde{\psi}_{g}^{t}|\hat{\mathbf{H}}| \tilde{\psi}_{g}^{t}\rangle$ relies on a semiclassical expansion of volume operator $\hat{V}_v$ \cite{Giesel:2006um}
\be
\hat{V}_v=\langle \hat{Q}_v\rangle^{2q}\lt[1+\sum_{n=1}^{2k+1}(-1)^{n+1}\frac{q(1-q)\cdots(n-1+q)}{n!}\lt(\frac{\hat{Q}_v^2}{\langle\hat{Q}_v\rangle^2}-1\rt)^n\rt],\quad q=1/4\label{expandvolume}
\ee
where $\langle \hat{Q}_v\rangle=\langle\psi^t_g|\hat{Q}_v|\psi^t_g\rangle$. This expansion is valid when $\langle \hat{Q}_v\rangle\gg \ell_p^6$.

We write $g_{i+1}(e)=g_i(e) [1+{\Delta\phi^a(e)\t^a}]$ where $\Delta\phi^a(e)$ parametrizes the infinitesimal change of $g(e)$ between two time steps. Eqs (\ref{eoms1}) and (\ref{eoms2}) reduce to follows (by using Lemma 4.1 in \cite{Han:2019vpw}): 
\be
- \frac{i a^2}{\kappa} {{M_1}^a}_b(g(e)) \frac{\Delta \bar{\phi}^b(e)}{\Delta \tau} &=& \frac{\partial }{\partial {\eps}^a(e)}\mathbf{H}\lt[g^\eps\rt] \Big|_{\vec{\eps}=0}\label{eom01}\\
- \frac{i a^2}{\kappa} {{M_2}^a}_b(g(e)) \frac{\Delta  {\phi}^b(e)}{\Delta \tau} &=& - \frac{\partial }{\partial {\bar{\eps}}^a(e)}{\bf H} \lt[g^\eps\rt]\Big|_{\vec{\eps}=0}\label{eom02}
\ee
where the left-hand sides become time derivatives as $\Delta\t\to0$, and
\be
M_{1}{}^a_{\ b}(g)&=&2\L^a_{\ c}(\vec{\theta})\L^{b}_{\ d}(\vec{\theta})\lt[\frac{p^c}{p}\frac{p^d}{p}-i\eps^{cde}p^e+\frac{p\cosh(p)}{\sinh(p)}\lt(\delta^{cd}-\frac{p^c}{p}\frac{p^d}{p}\rt)\rt],\\
M_{2}{}^a_{\ b}(g)&=&2\L^a_{\ c}(\vec{\theta})\L^{b}_{\ d}(\vec{\theta})\lt[\frac{p^c}{p}\frac{p^d}{p}+i\eps^{cde}p^e+\frac{p\cosh(p)}{\sinh(p)}\lt(\delta^{cd}-\frac{p^c}{p}\frac{p^d}{p}\rt)\rt], 
\ee
where $e^{\theta^{c} \tau^{c} / 2} \tau^{a} e^{-\theta^{c} \tau^{c} / 2}=\Lambda_{\ b}^{a}(\bm{\theta}) \tau^{b}$. The matrices $M_{1}{}^a_{\ b}(g)$ and $M_{2}{}^a_{\ b}(g)$ are nondegenerate since
\be
\det\lt(M_{1,2}(g)\rt)=\frac{\sinh^2(p)}{p^2}\neq 0.
\ee

We can write $\Delta\phi^a(e)$ as a linear combination of $\Delta p^a(e) = p_{i+1}^a(e) - p_i^a(e)  $ and $\Delta \theta^a(e) = \theta_{i+1}^a(e) - \theta_i^a(e) $
\be
	{\Delta  {\phi}}^a(e_I) = - \frac{1}{2}\Tr( g_{i}^{-1}(e) g_{i+1}(e) \tau^a ) = {{J_1}^a}_b(e)  \Delta p^a(e)+{{J_2}^a}_b(e)  \Delta \theta^a(e). 
\ee
at leading orders of $\Delta p^a(e)$ and $\Delta \theta^a(e)$. The holomorphic deformation $\eps^a(e)$ has the similar expression
\be
	{\eps}^a(e)  = - \frac{1}{2} \Tr( g^{-1}(e)  {g}^{\eps}(e)  \tau^a ) = {J_1}^a_{\ b}(e)  \delta p^a(e) + {J_2}^a_{\ b}(e)  \delta \theta^a(e)
\ee
where $\delta p^a(e)  $ and $\delta \theta^a(e)$ relates to $g^\eps(e)$ by
\be
g^\eps(e)=e^{-i \lt[p^a(e)+\delta p^a(e)\rt]\t^a/2}e^{\lt[\theta^a(e)+\delta \theta^a(e)\rt]\t^a/2}.
\ee 
$J_1,J_2$ are $3 \text{-by-} 3$ complex matrices whose elements depend on $p^a(e)$ and $\theta^a(e)$. We define $6\times 6$ matrices $J$ and $\tilde{J}$ as:
\be
	J =\left( \begin{array}{ll} J_1 & J_2 \\ \Bar{J}_1 & \Bar{J_2} \end{array} \right)\;, \qquad \tilde{J} =\left( \begin{array}{ll}  \Bar{J}_1 & \Bar{J_2} \\ J_1 & J_2 \end{array} \right)\;.
\ee
$J$ and $\tilde{J}$ satisfy
\be
   \left( \begin{array}{l} \bm{\eps}(e) \\ \bm{\bar{\eps}}(e) \end{array} \right) &=& J \left( \begin{array}{l}  {\delta {\bm p}(e)} \\ {\delta \bm{\theta}(e)} \end{array} \right) = \left( \begin{array}{ll} J_1 & J_2 \\ \Bar{J}_1 & \Bar{J_2} \end{array} \right) \left( \begin{array}{l}  {\delta {\bm p}(e)} \\ {\delta \bm{\theta}(e)} \end{array} \right), \\
  \left( \begin{array}{l} \Delta \bm{\bar{\phi}}(e) \\ \Delta \bm{\phi} (e) \end{array} \right) &=& \tilde{J} \left( \begin{array}{l}  \Delta {\bm p}(e) \\ \Delta \bm{\theta} (e) \end{array} \right) = \left( \begin{array}{ll}  \Bar{J}_1 & \Bar{J_2} \\ J_1 & J_2 \end{array} \right) \left( \begin{array}{l}  \Delta {\bm p}(e) \\ \Delta \bm{\theta} (e) \end{array} \right)
\ee
Here the bold letters $\bm{p},\bm{\theta}$ denotes the $3$-vectors $p^a,\theta^a$. Using above matrices Eqs.(\ref{eom01})  and (\ref{eom02}) becomes 
\be
  {T} \lt({\bm p},{\bm \theta}\rt) \left( \begin{array}{l}  {\Delta {\bm p}}(e)/{\Delta \tau} \\ {\Delta \bm{\theta}}(e)/{\Delta \tau}  \end{array} \right)  =  \frac{i\kappa}{a^2}\left( \begin{array}{l}  {\partial {\bf H} }/{\partial {\bm p} (e)} \\ {\partial {\bf H} }/{\partial \bm{\theta} (e)} \end{array} \right),\label{eom0n}
\ee
where
\be
{T}\lt({\bm p},{\bm \theta}\rt)= \left( \begin{array}{ll} J_1 & J_2 \\ \Bar{J}_1 & \Bar{J_2} \end{array} \right)^{T}\left( \begin{array}{ll} M_1 &\ \ 0 \\ \ 0 & - M_2 \end{array} \right)  \left( \begin{array}{ll}  \Bar{J}_1 & \Bar{J_2} \\ J_1 & J_2 \end{array} \right).
\ee
It is much more convenient to compute the right-hand side of Eq.\Ref{eom0n} than right-hand sides of Eqs.\Ref{eom01} and \Ref{eom02}, since ${\bf H}$ is expressed in terms of holonomies and fluxes. 

By the time continuous limit $\Delta\t\to0$, ${\Delta {\bm p}}(e)/{\Delta \tau}\to {\rmd {\bm p}}(e)/{\rmd \tau}$ and ${\Delta {\bm \theta}}(e)/{\Delta \tau}\to {\rmd {\bm \theta}}(e)/{\rmd \tau}$, so the semiclassical EOMs reduce to
\be
{T}\lt({\bm p},{\bm \theta}\rt) \left( \begin{array}{l}  {\rmd {\bm p}}(e)/{\rmd \tau} \\ {\rmd \bm{\theta}}(e)/{\rmd \tau}  \end{array} \right)  = \frac{i\kappa}{a^2} \left( \begin{array}{l}  {\partial {\bf H} }/{\partial {\bm p} (e)} \\ {\partial {\bf H} }/{\partial \bm{\theta} (e)} \end{array} \right) .\label{eom0}
\ee

The above computation is carried out in Mathematica. The matrix elements of $J$, $\tilde{J}$, and $T$ are lengthy. Their explicit formulae are given in \cite{github}.

As seen from Eq.\Ref{eom0}, the approximation $g(\t)$ of any solution $g_i$ of Eqs.\Ref{eoms1} and \Ref{eoms2} is not only continuous in $\t$ but also differentiable. Indeed, if a solution $g_i\simeq g(\t)$ failed to be differentiable, left-hand sides of Eq.\Ref{eom0} or Eqs.\Ref{eoms1} and \Ref{eoms2} would have blew up with small $\Delta\t$ and contradicted the finiteness of right-hand sides, i.e. $g_i$ could not be a solution.

\section{Semiclassical Dynamics as Hamiltonian Evolution}\label{Semiclassical Dynamics as Hamiltonian Evolution}

\subsection{Holonomy-Flux Poisson Algebra}

Since the semiclassical EOMs are expressed in terms of variables $p^a(e), \theta^a(e)$, it is useful to compute the Poisson algebra of $p^a(e), \theta^a(e)$ from the holonomy-flux algebra Eqs.\Ref{handh} - \Ref{pandp} by the relation $h(e)=e^{\theta^a(e)\t^a/2}$. The computation can be proceed as the following: We write Eq.\Ref{pandtheta} (at $e'=e$) as
\be
\left\{p^{a}(e), \theta^b(e)\right\}\frac{\partial h_{AB}(e)}{\partial\theta^b(e)} &=&\frac{\kappa}{a^{2}} \lt[\frac{\tau^{a}}{2} h\left(e\right)\rt]_{AB}.\label{hAB}
\ee
Among 4 matrix elements $h_{AB}(e)$, there are only 3 independent $h_{11}(e),h_{12}(e),h_{21}(e)$. The above equations with $AB=11,12,21$ form a matrix equation of three $3\times3$ matrices $U,\ V$, and $W$:
\be
U^a_{\ b}V^b_{\ AB}=\frac{\kappa}{a^{2}}W^a_{\ AB},\quad \text{where}\quad U^a_{\ b}=\left\{p^{a}(e), \theta^b(e)\right\},\quad V^b_{\ AB}=\frac{\partial h_{AB}(e)}{\partial\theta^b(e)},\quad W^a_{\ AB}=\lt[\frac{\tau^{b}}{2} h\left(e\right)\rt]_{AB}
\ee
where $AB=11,12,21$. Solving $U=\frac{\kappa}{a^{2}}W V^{-1}$ gives the following result:
\be
&&\left\{p^{a}(e), \theta^b(e)\right\}\equiv U^a_{\ b}({\bm \theta})\\
&=&\left(
\begin{array}{ccc}
 \frac{2 \theta _1^2+\theta  \left(\theta _2^2+\theta _3^2\right) \cot \left(\frac{\theta }{2}\right)}{2 \theta ^2} &
   -\frac{\theta _3^3+\left(\theta _1^2+\theta _2^2\right) \theta _3+\theta _1 \theta _2 \left(\theta  \cot
   \left(\frac{\theta }{2}\right)-2\right)}{2 \theta ^2} & \frac{1}{2} \left(\frac{\theta _1 \theta _3 \left(2-\theta
    \cot \left(\frac{\theta }{2}\right)\right)}{\theta ^2}+\theta _2\right) \\
 \frac{\theta _3^3+\left(\theta _1^2+\theta _2^2\right) \theta _3+\theta _1 \theta _2 \left(2-\theta  \cot
   \left(\frac{\theta }{2}\right)\right)}{2 \theta ^2} & \frac{2 \theta _2^2+\theta  \left(\theta _1^2+\theta
   _3^2\right) \cot \left(\frac{\theta }{2}\right)}{2 \theta ^2} & \frac{1}{2} \left(\frac{\theta _2 \theta _3
   \left(2-\theta  \cot \left(\frac{\theta }{2}\right)\right)}{\theta ^2}-\theta _1\right) \\
 -\frac{\theta _2 \theta _1^2+\theta _2 \left(\theta _2^2+\theta _3^2\right)+\theta _3 \theta _1 \left(\theta  \cot
   \left(\frac{\theta }{2}\right)-2\right)}{2 \theta ^2} & \frac{\theta _1^3+\left(\theta _2^2+\theta _3^2\right)
   \theta _1+\theta _2 \theta _3 \left(2-\theta  \cot \left(\frac{\theta }{2}\right)\right)}{2 \theta ^2} & \frac{2
   \theta _3^2+\theta  \left(\theta _1^2+\theta _2^2\right) \cot \left(\frac{\theta }{2}\right)}{2 \theta ^2} \\
\end{array}
\right)\nonumber
\ee
where $\theta_a\equiv\theta^a(e)$ and $\theta=\sqrt{\theta^a(e)\theta^a(e)}$. With this result we check that Eq.\Ref{hAB} with $AB=21$ is satisfied automatically.

The holonomy-flux algebra Eqs.\Ref{handh} - \Ref{pandp} implies the following Poisson algebra between $p^a(e)$ and $\theta^a(e)$
\be
\left\{\theta^a(e), \theta^b\left(e^{\prime}\right)\right\} &=&0 ,\label{pth1}\\
\left\{p^{a}(e), \theta^b\left(e^{\prime}\right)\right\} &=&\frac{\kappa}{a^{2}} \delta_{e, e^{\prime}} U^a_{\ b}({\bm \theta}) ,\label{pth2}\\
\left\{p^{a}(e), p^{b}\left(e^{\prime}\right)\right\} &=&-\frac{\kappa}{a^{2}} \delta_{e, e^{\prime}} \varepsilon_{a b c} p^{c}\left(e^{\prime}\right),\label{pth3}
\ee

A straight-forward computation demonstrates that Eqs.\Ref{pth1} - \Ref{pth3} implies the holonomy-flux algebra Eqs.\Ref{handh} - \Ref{pandp}. Thus the holonomy-flux algebra and the Poisson algebra between $p^a(e)$ and $\theta^a(e)$ in Eqs.\Ref{pth1} - \Ref{pth3} are equivalent.

\subsection{Hamilton's equations}

We would like to relate EOMs \Ref{eom0} to Hamilton's equations with the discrete physical Hamiltonian ${\bf H}$ and symplectic structure of holonomy-flux algebra. Firstly
\be
\lt\{p^a(e), {\bf H}\rt\}=\lt\{p^a(e),p^b(e)\rt\}\frac{\partial {\bf H}}{\partial p^b(e)}+\lt\{p^a(e),\theta^b(e)\rt\}\frac{\partial {\bf H}}{\partial \theta^b(e)}\nonumber\\
\lt\{\theta^a(e), {\bf H}\rt\}=\lt\{\theta^a(e),p^b(e)\rt\}\frac{\partial {\bf H}}{\partial p^b(e)}+\lt\{\theta^a(e),\theta^b(e)\rt\}\frac{\partial {\bf H}}{\partial \theta^b(e)}.
\ee
We define the matrix
\be
P({\bm p},{\bm \theta})=\lt(\begin{array}{cc}
\lt\{p^a(e),p^b(e)\rt\} & \lt\{p^a(e),\theta^b(e)\rt\}\\
\lt\{\theta^a(e),p^b(e)\rt\} & 0
\end{array}\rt).
\ee
Applying $P$ to the EOMs \Ref{eom0} gives
\be
-\frac{ia^2}{\kappa} P\lt({\bm p},{\bm \theta}\rt){T}\lt({\bm p},{\bm \theta}\rt) \left( \begin{array}{l}  {\rmd {\bm p}}(e)/{\rmd \tau} \\ {\rmd \bm{\theta}}(e)/{\rmd \tau}  \end{array} \right)  = \left( \begin{array}{l}  \lt\{\bm{p}(e), {\bf H}\rt\} \\ \lt\{\bm{\theta}(e), {\bf H}\rt\} \end{array} \right).
\ee
By using the explicit formula of ${T}\lt({\bm p},{\bm \theta}\rt)$ and Poisson brackets in $P\lt({\bm p},{\bm \theta}\rt)$, we obtain the following simple result
\be
-\frac{ia^2}{\kappa} P\lt({\bm p},{\bm \theta}\rt){T}\lt({\bm p},{\bm \theta}\rt)=1_{6\times6}.\label{PT1}
\ee
This shows that the semiclassical EOMs from the path integral is equivalent to Hamilton's equations with the discrete physical Hamiltonian ${\bf H}$:
\be
\frac{\rmd {p}^a(e)}{\rmd \tau}=\lt\{{p}^a(e),\ {\bf H}\rt\}, \quad \frac{\rmd {\theta}^a(e)}{\rmd \tau}=\lt\{ {\theta}^a(e),\ {\bf H}\rt\},\label{hamiton4.11}
\ee
where the Poisson brackets are given by Eqs.\Ref{pth1} - \Ref{pth3}, or equivalently, by the holonomy-flux algebra Eqs.\Ref{handh} - \Ref{pandp}. In general, the time evolution of any phase space function $f( {p}^a(e), {\theta}^a(e))$ or $f( {p}^a(e), h(e))$ is governed by
\be
\frac{\rmd f}{\rmd \tau}=\lt\{f, \ {\bf H}\rt\}.\label{hamiton}
\ee

Mathematica is employed for all above computations, including computing $\{p^a(e),\theta^b(e)\}$, check the equivalence between Eqs.\Ref{pth1} - \Ref{pth3} and holonomy-flux algebra, and verifying Eq.\Ref{PT1}. The Mathematica files can be found in \cite{github}.

Moreover the closure condition \Ref{closure0} is equivalent to $\sum_{I=1}^3\sum_{s=\pm}p^a(e_{v;I,s})=0$. The Hamiltonian flow generated by $G^a_v:=\sum_{I=1}^3\sum_{s=\pm}p^a(e_{v;I,s})$ in a $\cp_\g$ is SU(2) gauge transformation. Since ${\bf H}$ is SU(2) gauge invariant,
\be
\frac{\rmd G^a_v}{\rmd \t}=\lt\{G^a_v,\,{\bf H}\rt\}=0.
\ee
So the closure condition \Ref{closure0} is preserved in the time evolution. Given a solution $p^a(\t,e),\theta^b(\t,e)$ satisfying Eq.\Ref{hamiton}, its gauge transformation still satisfies Eq.\Ref{hamiton}:
\be
\lt\{\{f,\,G^a_v\},\,{\bf H}\rt\}&=&-\lt\{\{G^a_v,\,{\bf H}\},\,f\rt\}-\lt\{\{{\bf H},\,f\},\,G^a_v\rt\}=\lt\{f,\,\frac{\rmd G^a_v}{\rmd \t}\rt\}+\lt\{\frac{\rmd f}{\rmd \tau},\,G^a_v\rt\}\nonumber\\
&=&\frac{\rmd }{\rmd \tau}\{f,\,G^a_v\}.
\ee
Recall that the initial state in Eq.\Ref{Agg} is labelled by the gauge equivalence class $[g']$, the trajectory in the reduced phase space determined by the Hamiltonian flow \Ref{hamiton} is unique up to SU(2) gauge transformations, in the phase space regime where ${\bf H}$ is a smooth function in $p^a,\theta^a$. 

Note that due to the absolute-value and square-root in ${\bf H}$, ${\bf H}$ is non-differentiable at $C_v^2-\frac{\a}{4}\sum_{a=1}^3 C_{a,v}^2=0$, at which the uniqueness of solution cannot be established. As it is discussed in Section \ref{Lattice Continuum Limit}, these irregularities are avoided if initial states $\Psi_{[g']}^t$ are semiclassical in the sense that $[g']$ is in the classically allowed regime of the phase space. The classically allowed regime satisfies non-holonomic constraints required by the classical gravity-dust theory.

\section{Action Principle}\label{Action Principle}

Here we present another routine to derive the classical EOMs (the Hamilton's equation \Ref{hamiton}). We are going to firstly take the time continuous limit of the discrete action $S[g,h]$, then derive EOMs, in contrast to the above procedure in which discrete EOMs are derived firstly from the path integral, then take the time continuous limit.

Recall $S[g,h]$ in Eq.\Ref{Sgh}, we write
\be
g_i=g(\t),\quad g_{i+1}=g(\t+\Delta\t),
\ee
and expand summands in $S[g,h]$ in $\Delta\t$:
\be
&&\frac{\langle\psi_{g_{i+1}}^{t}|\hat{\mathbf{H}}| \psi_{g_{i}}^{t}\rangle}{\langle\psi_{g_{i+1}}^{i} | \psi_{g_{i}}^{t}\rangle}+i \tilde{\varepsilon}_{i+1, i}\left(\frac{\Delta \tau}{\hbar}\right)= \langle\psi_{g(\t)}^{t}|\hat{\mathbf{H}}| \psi_{g(\t)}^{t}\rangle+O(\Delta\t),\\
&&K\left(g_{i+1}, g_{i}\right)=\Delta\t\sum_{e\in E(\g)} i G_{ab}\big(\bm{\theta}(\t,e)\big)p^a(\t,e)\frac{\rmd \theta^b(\t,e)}{\rmd\t}+O(\Delta\t^2).
\ee
The $3\times3$ real matrix $G_{ab}\big(\bm{\theta}\big)$ is given by
\be
\left(
\begin{array}{ccc}
 -\frac{\left(\theta  \theta _1^2+\left(\theta _2^2+\theta _3^2\right) \sin (\theta )\right)}{\theta ^3} & -\frac{   \left(\theta _1 \theta _2 (\theta -\sin (\theta ))+\theta  \theta _3 (\cos (\theta )-1)\right)}{\theta ^3} & \frac{   \left(\theta _1 \theta _3 (\sin (\theta )-\theta )+\theta  \theta _2 (\cos (\theta )-1)\right)}{\theta ^3} \\
 \frac{\theta  \theta _3 (\cos (\theta )-1)-\theta _1 \theta _2 (\theta -\sin (\theta ))}{\theta ^3} & -\frac{   \left(\theta  \theta _2^2+\left(\theta _1^2+\theta _3^2\right) \sin (\theta )\right)}{\theta ^3} & -\frac{\left(\theta
   _2 \theta _3 (\theta -\sin (\theta ))+\theta  \theta _1 (\cos (\theta )-1)\right)}{\theta ^3} \\
 -\frac{\left(\theta _1 \theta _3 (\theta -\sin (\theta ))+\theta  \theta _2 (\cos (\theta )-1)\right)}{\theta ^3} &
   \frac{\left(\theta _2 \theta _3 (\sin (\theta )-\theta )+\theta  \theta _1 (\cos (\theta )-1)\right)}{\theta ^3} &
   -\frac{\left(\theta  \theta _3^2+\left(\theta _1^2+\theta _2^2\right) \sin (\theta )\right)}{\theta ^3} \\
\end{array}
\right)
\ee
where $\theta_a\equiv\theta^a(e)$ and $\theta=\sqrt{\theta^a(e)\theta^a(e)}$. 

We find that $G_{ab}({\bm \theta})$ closely relates to $U^a_{\ b}({\bm \theta})=\left\{p^{a}(e), \theta^b(e)\right\}$ by
\be
G({\bm \theta})^T U({\bm \theta})=U({\bm \theta}) G({\bm \theta})^T=-\frac{\kappa}{a^2}\, 1_{3\times3}.
\ee
We define new variables 
\be
X^b(\t,e)= G_{ab}\big(\bm{\theta}(\t,e)\big)p^a(\t,e)
\ee
and interestingly, we obtain the following result:

\begin{Theorem} 

The following (equal-time) Poisson algebra between $X^a$ and $\theta^a$ is equivalent to the holonomy-flux algebra
\be
\lt\{X^a(e),\theta^b(e')\rt\}=-\frac{\kappa}{a^2}\delta^{ab}\delta_{e,e'},\quad \lt\{X^a(e),X^b(e')\rt\}= \lt\{\theta^a(e),\theta^b(e')\rt\}=0.\label{Xth}
\ee
$X^a(e), \theta^a(e)$ form local Darboux coordinate on the reduced phase space of LQG.
\end{Theorem}

\textbf{Proof:} The first relation is equivalent to Eq.\Ref{pth1}
\be
\lt\{X^a(e),\theta^b(e')\rt\}= G_{ca}\big(\bm{\theta}(e)\big)\left\{p^{c}(e), \theta^b(e')\right\}= G_{ca}\big(\bm{\theta}(e)\big)U^c_{\ b}({\bm \theta}(e))\delta_{e,e'}=-\frac{\kappa}{a^2}\delta^{ab}\delta_{e,e'}.
\ee 
Secondly, 
\be
&&\lt\{X^a(e),X^b(e')\rt\}=\lt\{G_{ca}\big(\bm{\theta}(e)\big)p^c(e),G_{db}\big(\bm{\theta}(e')\big)p^d(e')\rt\}\nonumber\\
&=&G_{ca}\big(\bm{\theta}(e)\big)G_{db}\big(\bm{\theta}(e')\big)\lt\{p^c(e),p^d(e')\rt\}- G_{db}\big(\bm{\theta}(e')\big)p^c(e)\frac{\partial G_{ca}\big(\bm{\theta}(e)\big)}{\partial\theta^f(e)}\lt\{p^d(e'),\theta^f(e)\rt\}\nonumber\\
&&+\ G_{ca}\big(\bm{\theta}(e)\big)p^d(e')\frac{\partial G_{db}\big(\bm{\theta}(e')\big)}{\partial \theta^f(e')}\lt\{p^c(e),\theta^f(e')\rt\}\nonumber\\
&=&G_{ca}\big(\bm{\theta}(e)\big)G_{db}\big(\bm{\theta}(e')\big)\lt\{p^c(e),p^d(e')\rt\} + \frac{\kappa}{a^2}\delta_{e,e'} p^c(e)\lt[\frac{\partial G_{ca}\big(\bm{\theta}(e)\big)}{\partial\theta^b(e)}-\frac{\partial G_{cb}\big(\bm{\theta}(e)\big)}{\partial \theta^a(e)}\rt]
\ee
is vanishing because
\be
\lt\{p^c(e),p^d(e)\rt\}=-\frac{\kappa}{a^2}G^{-1}_{ac}\big(\bm{\theta}(e)\big)G^{-1}_{bd}\big(\bm{\theta}(e')\big)\lt[\frac{\partial G_{ea}\big(\bm{\theta}(e)\big)}{\partial\theta^b(e)}
-\frac{\partial G_{eb}\big(\bm{\theta}(e)\big)}{\partial \theta^a(e)}\rt]p^e(e),
\ee
which can be checked straight-forwardly. The Mathemaica file for the above computation is provided in \cite{github}.

$\Box$

Although the Poisson algebra Eq.\Ref{Xth} is simple, SU(2) gauge transformations of $X^a(e),\theta^a(e)$ are complicated. In contrast, the holonomy-flux algebra uses variables $p^a(e), h(e)$ that have simple SU(2) gauge transformations, but sacrifices the simplicity of Poisson brackets.

As a result we obtain the following time continuous limit $\cs[g,h]=\lim_{\Delta\t\to0}S[g,h]$
\be
\cs[g,h]&=& i \int_0^T\rmd\t \lt[\sum_{e\in E(\g)}X^a(\t,e)\frac{\rmd \theta^a(\t,e)}{\rmd \t}-\frac{\kappa}{a^2} \langle\psi_{g(\t)}^{t}|\hat{\mathbf{H}}| \psi_{g(\t)}^{t}\rangle\rt]\nonumber\\
&=&i \int_0^T\rmd\t \lt[\sum_{e\in E(\g)}X^a(\t,e)\frac{\rmd \theta^a(\t,e)}{\rmd \t}-\frac{\kappa}{a^2} \Big({\mathbf{H}}\lt[{\bm p}(\t),{\bm \theta}(\t)\rt]+O(\hbar)\Big)\rt]
\ee 
where $\langle\psi_{g(\t)}^{t}|\hat{\mathbf{H}}| \psi_{g(\t)}^{t}\rangle={\mathbf{H}}\lt[{\bm p}(\t),{\bm \theta}(\t)\rt]+O(\hbar)$.

The Poisson algebra Eq.\Ref{Xth}, or equivalently the holonomy-flux algebra, can be obtained from the above $\cs[g, h]$ by the Legendre transformation. $\cs[g, h]$ provides an action principle for the LQG (reduced) phase space and the quantization.

By the time continuous limit, the path integral formula \Ref{Agg} becomes a standard phase space path integral
\be
\int \lt[DXD\theta\rt]\, \mu[X,\theta]\, e^{\frac{i}{t}\int_0^T\rmd\t \lt[\sum_{e\in E(\g)}X^a(\t,e)\frac{\rmd \theta^a(\t,e)}{\rmd \t}-\frac{i\kappa}{a^2} \Big({\mathbf{H}}+O(\hbar)\Big)\rt]}
\ee
up to $O(\hbar)$ in the action and a measure factor $\mu[X,\theta]$ (containing $\nu[g]$ and the Jacobian for transforming $\rmd g\to \rmd X\rmd\theta$). The path integral formula becomes an infinite dimension integral, thus may be mathematically ill-defined. This path integral relates to a starting point in \cite{link,Han:2009bb}.

The variational principle $\delta \cs=0$ gives the Hamilton's equation (up to $O(\hbar)$)
\be
\frac{\rmd \theta^a(e)}{\rmd \t}=\frac{\kappa}{a^2} \frac{\partial\mathbf{H}}{\partial X^a(e)},\quad \frac{\rmd X^a(e)}{\rmd \t}=-\frac{\kappa}{a^2} \frac{\partial\mathbf{H}}{\partial \theta^a(e)}
\ee
For any phase space function $f({\bm X},{\bm \theta})$, its time evolution is given by
\be
\frac{\rmd f}{\rmd \tau}=\lt\{f, \ {\bf H}\rt\},
\ee
which is identical to Eq.\Ref{hamiton}. It shows that the time continuous limit and variational principle are commutative when acting on $S[g,h]$.

\section{Lattice Continuum Limit}\label{Lattice Continuum Limit}

In this section, we demonstrate the relation between the semiclassical EOMs \Ref{eom0} (or equivalently \Ref{hamiton}) from path integral and classical reduced phase space EOMs \Ref{hamitoncon0} of gravity-dust system in the continuum. We are going to take the continuum limit of the cubic lattice $\g$, i.e. send the total number $|V(\g)|$ of vertices to infinity, and show that \Ref{eom0} recovers \Ref{hamitoncon0} in this limit. Defining $\mu\sim |V(\g)|^{-3}$ to be the coordinate length of every lattice edge, the lattice continuum limit is given by $\mu\to0$. More precisely, recall that semiclassical EOMs are derived with $t=\ell_P^2/a^2\to0$ and $\langle \hat{Q}_v\rangle\sim \mu^6\gg \ell_P^6$ (see Eq.\Ref{expandvolume}), the lattice continuum limit takes us to the regime 
\be
\ell_P\ll\mu\ll a,
\ee
where $a$ is a macroscopic unit, e.g. $a=1mm$. When keeping $a$ fixed, the lattice continuum limit sends $\mu\to0$ after the semiclassical limit $\ell_P\to0$ (from which EOMs are derived) so $\ell_P\ll\mu$ is kept.

We rescale $\theta^a(e),p^a(e)$:
\be
\theta^a\lt(e_I(v)\rt)=\mu \Fa^a_I(v),\quad p^a\lt(e_I(v)\rt)=\frac{2\mu^2}{\b a^2} \Fe_a^I(v),\label{thApE}
\ee   
where $\Fa^a_I(v), \Fe_a^I(v)$ behave as follows in the lattice continuum limit:
\be
\Fa^a_I(v)=A^a_I(v)+O(\mu),\quad \Fe_a^I(v)=E_a^I(v)+O(\mu).\label{elementcon}
\ee
Here $A^a_I(v)=A^a_j(v)\dot{e}_I(v)^j$ and $E_a^I(v)=E_a^j(v)\dot{e}_I(v)^j$ are smooth fields $({\bm A},{\bm E})$ evaluated at the vertex $v$. $\dot{e}_I(v)$ is the tangent vector of $e_I(v)$ at $v$. $A^a_I(v), E_a^I(v)$ are coordinate components of $({\bm A},{\bm E})$ when we take $\dot{e}_I(v)\equiv \partial/\partial \sig^I$ ($I=1,2,3$) as coordinate basis. $\sig^I$ is such that the coordinate length of $e_I(v)$ is $\mu$.


Inserting the $\mu$-expansion of $\theta^a(e),p^a(e)$ in $T({\bm p},{\bm \theta})$ of Eq.\Ref{eom0} gives:
\be
T({\bm p},{\bm \theta})=\left(
\begin{array}{cccccc}
 0 & 0 & 0 & -i & 0 & 0 \\
 0 & 0 & 0 & 0 & -i & 0 \\
 0 & 0 & 0 & 0 & 0 & -i \\
 i & 0 & 0 & 0 & 0 & 0 \\
 0 & i & 0 & 0 & 0 & 0 \\
 0 & 0 & i & 0 & 0 & 0 \\
\end{array}
\right)+O(\mu).
\ee
So the left hand side of Eq.\Ref{eom0} becomes
\be
{T}\lt({\bm p},{\bm \theta}\rt) \left( \begin{array}{l}  \frac{\rmd {\bm p}(e_I(v))}{\rmd \tau} \\ \frac{\rmd \bm{\theta}(e_I(v))}{\rmd \tau}  \end{array} \right) =i\left( \begin{array}{l}  -\mu \frac{\rmd {\bm A}_I(v)}{\rmd \tau}+O(\mu^2) \\ \frac{2\mu^2}{\b a^2}\frac{\rmd \bm{E}^I(v)}{\rmd \tau} +O(\mu^3) \end{array} \right).\label{kincon}
\ee

On the right hand side of Eq.\Ref{eom0},
\be
\frac{\partial{\bf H}[\bm{p},\bm{\theta}]}{\partial p^a(e_I(v))}
=\frac{\b a^2}{2\mu^2}\frac{\partial{\bf H}[\Fe,\Fa]}{\partial \Fe^I_a(v)},\quad 
\frac{\partial{\bf H}[\bm{p},\bm{\theta}]}{\partial \theta^a(e_I(v))}
=\frac{1}{\mu}\frac{\partial{\bf H}[\Fe,\Fa]}{\partial \Fa^a_I(v)}.
\label{dercon02}
\ee
${\bf H}[\Fe,\Fa]$ is obtained from ${\bf H}[\bm{p},\bm{\theta}]$ by changing variables \Ref{thApE}. 
Derivatives of ${\bf H}$ reduces to derivatives of $C_v$ and $C_{a,v}$:
\be
\frac{\partial{\bf H}}{\partial \Fe^I_a(v')}&=&\sum_{v\in V(\g)}s_v\lt[\frac{C_v}{H_v}\,\frac{\partial C_v}{\partial \Fe^I_a(v')}-\frac{\a}{4}\sum_{b=1}^3\frac{C_{b,v}}{H_v}\frac{\partial C_{b,v}}{\partial \Fe^I_a(v')}\rt],\label{Hder0000}\\
\frac{\partial{\bf H}}{\partial \Fa^a_I(v')}&=&\sum_{v\in V(\g)}s_v\lt[\frac{C_v}{H_v}\,\frac{\partial C_v}{\partial \Fa^a_I(v')}-\frac{\a}{4}\sum_{b=1}^3\frac{C_{b,v}}{H_v}\frac{\partial C_{b,v}}{\partial \Fa^a_I(v')}\rt],\label{Hder0001}
\ee
where $H_v=\sqrt{|C_v^2-\frac{\a}{4}\sum_{a=1}^3C_{a,v}^2|}$ and $s_v=\mathrm{sgn}(C_v^2-\frac{\a}{4}\sum_{a=1}^3C_{a,v}^2)$. We have assumed that variations of $\Fe^I_a(v')$ and $\Fa^a_I(v')$ (for computing above derivatives) do not make any $s_v$ jump, so derivatives of $s_v$ are zero. Without this assumption, Hamilton's equations \Ref{hamiton4.11} is ill-defined because ${\bf H}$ is not differentiable as $s_v$ jumps. Semiclassial EOMs are singular at $C_v^2-\frac{\a}{4}\sum_{a=1}^3C_{a,v}^2=0$.

Computing explicitly Poisson brackets $h(e)\{h(e)^{-1},V_v\}$ and $h(e)\{h(e)^{-1},K\}$ makes
$C_v$ and $C_{a,v}$ as polynomials generated by following quantities
\be
h(e_I(v))&=&e^{\mu \Fa_I^a(v)},\quad
p^a(e_I(v))=\frac{2\mu^2}{\b a^2} \Fe_a^I(v),\label{interffff1}\\
{Q}_v ^{-\half}&=&\mu^{-3}\fq(v)^{-\half},\quad \fq(v)=\frac{1}{6}\eps_{IJK}\eps^{abc}\Fe_a^I(v)\Fe_b^J(v)\Fe_c^K(v),\label{interffff2}
\ee
where $Q_v$ is the classical limit of $\hat{Q}_v$ in Eq.\Ref{Qv}.

In the following we often use the short-hand notation 
\be
\ff_\a(v)&=&\lt(\Fe_a^I(v),\, \Fa_I^a(v),\, \fq(v)^{-\half}\rt)=f_\a(v)+O(\mu),\nonumber\\
f_\a(v)&=&\lt(E_a^I(v),\, A_I^a(v),\, q(v)^{-\half}\rt),\quad q(v)=\frac{1}{6}\eps_{IJK}\eps^{abc}E_a^I(v)E_b^J(v)E_c^K(v).
\ee

We apply Eqs.\Ref{interffff1} and \Ref{interffff2} to $C_v$ and $C_{a,v}$ and expand in $\mu$ (but do not recover smooth fields $f_\a$ from $\ff_\a$). $C_v\ \text{and}\ C_{a,v}$ can be cast into the following pattern (see Appendix \ref{typeexplanation} for an explanation):
\be
C_v\ \text{or}\ C_{a,v}
&=&\mu \sum_{\a,\b,J,K, N^\pm_J,M^\pm_K}F^{\a,\b}_{N^\pm_J,M^\pm_K}(\vec{v})\Delta_{J,N^\pm_J}\ff_\a(\tilde{v}_1)\,\Delta_{K,M^\pm_K}\ff_\b(\tilde{v}_2)\nonumber\\
&&+\ \mu^2 \sum_{\a,J, N^\pm_J}F^\a_{N^\pm_J}(\vec{v})\Delta_{J,N^\pm_J}\ff_\a(\tilde{v})+ \mu^3F(\vec{v})+O(\mu^4),\label{typical}
\ee
where
\be
\Delta_{J,N^\pm_J}\ff_\a(\tilde{v})=\ff_\a(\tilde{v}+N^+_J\mu\hat{J})-\ff_\a(\tilde{v}-N^-_J\mu\hat{J}).
\ee
$\vec{v}=(v_1,v_2,\cdots)$ and $\tilde{v},\tilde{v}_1,\tilde{v}_2$ are some vertices whose distance from $v$ are of $O(\mu)$. $-3\leq N^\pm_J\leq 3$ ($N_J^+\neq -N_J^-$) are integers and $\hat{J}$ is the lattice vector along the $J$-th direction. Nonzero $N^\pm_J$ reflect correlations among variables at neighboring vertices in $C_v$ and $C_{a,v}$. Correlations are not only among nearest neighbors. $F^\a_{N^+_J,N^-_J}(\vec{v})$ and $F(\vec{v})$ (with $\vec{v}=(v_1,v_2,\cdots)$ a finite sequence of vertices $v_i$) are polynomials of $\ff_\a({v}_i)$ where ${v}_i=v+\sum_J N_i(J)\mu\hat{J}_i$ ($J_i\in\{1,2,3\}$ and integer $N_i\in[-3, 3]$) are vertices at or near $v$. Parameters $\a,\ \b$, $N^\pm,\ M^\pm$, $J$, $\vec{v}$, and $\tilde{v},\tilde{v}_1,\tilde{v}_2$ are determined by patterns of variables and Poisson brackets in $C_v$,$C_{a,v}$, thus are independent of $v$.

If $f_\a(v)$ evaluate as smooth fields at lattice vertex $v$, the continuum limit of \Ref{typical} is of $O(\mu^3)$:
\be
C_v\ \text{or}\ C_{a,v}&=&\mu^3\sum_{\a,\b,J,K}\lt[\sum_{ N^\pm_J,M^\pm_K}(N^+_J+N^-_J)(M^+_K+M^-_K)\cf^{\a,\b}_{N^\pm_J,M^\pm_K}({v})\rt]\partial_Jf_\a(v)\partial_Kf_\b(v)\nonumber\\
&&+\ \mu^3\sum_{\a,J}\lt[\sum_{ N^\pm_J}(N^+_J+N^-_J)\cf^\a_{N^\pm_J}({v})\rt]\partial_J f_\a(v)+\mu^3\cf({v})+O(\mu^4).\label{typicalcon}
\ee
$\cf^{\a,\b}_{N^\pm_J,M^\pm_K}({v}),\ \cf^\a_{N^+_J,N^-_J}({v})$, and $\cf({v})$ are continuum limit of $F^{\a,\b}_{N^\pm_J,M^\pm_K}(\vec{v}),\ F^\a_{N^+_J,N^-_J}(\vec{v})$ and $F(\vec{v})$:
\be
F^{\a,\b}_{N^\pm_J,M^\pm_K}(\vec{v})&=&\cf^{\a,\b}_{N^\pm_J,M^\pm_K}({v})+O(\mu),\nonumber\\
F^\a_{N^+_J,N^-_J}(\vec{v})&=&\cf^\a_{N^+_J,N^-_J}({v})+O(\mu),\nonumber\\ 
F(\vec{v})&=&\cf({v})+O(\mu).
\ee 
They are given by $F^{\a,\b}_{N^\pm_J,M^\pm_K}(\vec{v}),\ F^\a_{N^+_J,N^-_J}(\vec{v})$ and $F(\vec{v})$ with all $v_i\to v$ and applying Eq.\Ref{elementcon}. $\cf^{\a,\b}_{N^\pm_J,M^\pm_K}({v}),\ \cf^\a_{N^+_J,N^-_J}({v})$ and $\cf({v})$ are polynomials of $E_a^I(v),\, A_I^a(v),\, q(v)^{-\half}$. Let's take an example for illustration,
\be
\fq(v_1)^{-\half}\Fe_1^2(v_2)\Fe_2^1(v_3) =q(v)^{-\half}E_1^2(v)E_2^1(v)+O(\mu).
\ee  
The leading term on the right hand side corresponds to a term in $\cf^{\a,\b}_{N^\pm_J,M^\pm_K}({v}),\ \cf^\a_{N^+_J,N^-_J}({v})$ or $\cf({v})$.

We check that $C_v,\ C_{a,v}$, ${\bf H}$, and $G^a_v$ have correct continuum limits (i.e. \Ref{typicalcon} recovers continuum expressions of scalar and vector constraints $\cc(v), \cc_{a}(v)$ up to a prefactor $\mu^3$):
\be
C_v&=&\mu^3 \cc(v)+O(\mu^4),\label{Ccc}\\
C_{a,v}&=&\mu^3 \cc_{a}(v)+O(\mu^4),\label{Cjccj}\\
H_v&=&\mu^3 h(v)+O(\mu^4)=\mu^3 \sqrt{\lt|\cc(v)^2-\frac{\a}{4}\sum_{a=1}^3\cc_{a}(v)^2\rt|}+O(\mu^4)\\
{\bf H}&=&\sum_v\mu^3 \sqrt{\lt|\cc(v)^2-\frac{\a}{4}\sum_{a=1}^3\cc_{a}(v)^2\rt|}+O(\mu^4)\simeq \int_\cs\rmd^3\sig \sqrt{\lt|\cc(\sig)^2-\frac{\a}{4}\sum_{a=1}^3\cc_{a}(\sig)^2\rt|},\\
G^a_v&=&\frac{2\mu^3}{\b a^2} D_jE^j_a(v)+O(\mu^3).
\ee
Mathematica codes for deriving Eqs.\Ref{Ccc} and \Ref{Cjccj} are given in \cite{github}. The last relation indicates that the closure condition \Ref{closure0} reduces to the Gauss constraint in the lattice continuum limit. 

Continuum limit of $s_v$ is given by
\be
s_v=\mathrm{sgn}\lt(C_v^2-\frac{\a}{4}\sum_{a=1}^3C_{a,v}^2\rt)=\mathrm{sgn}\lt(\cc(v)^2-\frac{\a}{4}\sum_{a=1}^3\cc_{a}(v)^2+O(\mu)\rt).
\ee
$\cc,\cc_{a}$ are smooth fields in the continuum.

Given $v'\in V(\g)$, we assume $v'$ is inside a neighborhood $U\subset\cs$, such that $s_v=s_U$ is a constant for all $v\in U$ and the coordinate distance $r(v',\partial U)$ between $v'$ and any point in $\partial U$ satisfy $r(v',\partial U)\gg \mu$. This is an assumption for phase space points at which derivatives in Eqs.\Ref{Hder0000} and \Ref{Hder0001} are computed. This assumption is necessary for the lattice continuum limit of Eqs.\Ref{Hder0000} and \Ref{Hder0001}, because otherwise as $\mu\to0$, $v'$ approaches the boundary where $C_v^2-\frac{\a}{4}\sum_{a=1}^3C_{a,v}^2=0$, then $s_{v'}$ jumps by variations for computing derivatives of ${\bf H}$ thus invalidates Eqs.\Ref{Hder0000} and \Ref{Hder0001}.

We compute the following term in Eq.\Ref{Hder0000}:
\be
&&\sum_{v\in V(\g)} s_v\frac{C_v}{H_v}\frac{\partial C_v}{\partial \Fe^I_a(v')}\nonumber\\
&=&\mu s_U \sum_{v\in U} \frac{C_v}{H_v}\sum_{\a,\b,J,K, N^\pm_J,M^\pm_K}\sum_i\frac{\partial F^{\a,\b}_{N^\pm_J,M^\pm_K}(\vec{v})}{\partial \Fe^I_a(v_i)}\delta_{v',v_i}\Delta_{J,N^\pm_J}\ff_\a(\tilde{v}_1)\,\Delta_{K,M^\pm_K}\ff_\b(\tilde{v}_2)\nonumber\\
&&+\ \mu s_U \sum_{v\in U} \frac{C_v}{H_v}\sum_{\a,\b,J,K, N^\pm_J,M^\pm_K} F^{\a,\b}_{N^\pm_J,M^\pm_K}(\vec{v})\lt[\frac{\partial \ff_\a(v')}{\partial \Fe^I_a(v')}\delta_{v',\tilde{v}_1+N^+_J\mu\hat{J}}-\frac{\partial \ff_\a(v')}{\partial \Fe^I_a(v')}\delta_{v',\tilde{v}_1-N^-_J\mu\hat{J}}\rt]\Delta_{K,M^\pm_K}\ff_\b(\tilde{v}_2)\nonumber\\
&&+\ \mu s_U \sum_{v\in U} \frac{C_v}{H_v}\sum_{\a,\b,J,K, N^\pm_J,M^\pm_K} F^{\a,\b}_{N^\pm_J,M^\pm_K}(\vec{v})\Delta_{J,N^\pm_J}\ff_\a(\tilde{v}_1)\lt[\frac{\partial \ff_\b(v')}{\partial \Fe^I_a(v')}\delta_{v',\tilde{v}_2+M^+_K\mu\hat{K}}-\frac{\partial \ff_\b(v')}{\partial \Fe^I_a(v')}\delta_{v',\tilde{v}_2-M^-_K\mu\hat{K}}\rt]\nonumber\\
&&+\ \mu^2 s_U \sum_{v\in U} \frac{C_v}{H_v}\sum_{\a,J, N^\pm_J}\sum_i\frac{\partial F^\a_{N^\pm_J}(\vec{v})}{\partial \Fe^I_a(v_i)}\delta_{v',v_i}\Delta_{J,N^\pm_J}\ff_\a(\tilde{v})\nonumber\\
&&+\ \mu^2 s_U\sum_{v\in U} \frac{C_v}{H_v}\sum_{\a,J, N^\pm_J}F^\a_{N^\pm_J}(\vec{v})\lt[\frac{\partial \ff_\a(v')}{\partial \Fe^I_a(v')}\delta_{v',\tilde{v}+N^+_J\mu\hat{J}}-\frac{\partial \ff_\a(v')}{\partial \Fe^I_a(v')}\delta_{v',\tilde{v}-N^-_J\mu\hat{J}}\rt]\nonumber\\
&&+\ \mu^3 s_U\sum_{v\in U} \frac{C_v}{H_v}\sum_i\frac{\partial F(\vec{v})}{\partial \Fe^I_a(v_i)}\delta_{v',v_i}\nonumber\\
&&+\ O(\mu^4).\label{derivativeC}
\ee
Two sums $\sum_v$ and $\sum_{\a, \b,J,K, N^\pm_J,M^\pm_K}$ (or $\sum_{\a,J, N^\pm_J}$, $\sum_i$) can be interchanged since $\a,J, N^+_J,N^-_J,N_i$ are independent of $v$. Kronecker deltas in Eq.\Ref{derivativeC} are nonzero only if $v$ inside $U$ by the assumption $r(v',\partial U)\gg \mu$, since distances from $v_i,\tilde{v},\tilde{v}_{1,2}$ to $v$ is of $O(\mu)$. $\sum_{v\in U}$ in the result can be freely extend to $\sum_{v}$ over all $v\in V(\g)$, because $v$ outside $U$ has no contribution.

In the first term in the result of Eq.\Ref{derivativeC}, $\delta_{v_i,v'}$ restricts $v=v'-\delta_i$, where $\delta_i=v_i-v=\sum_J N_i(J)\mu \hat{J}_i$. We denote by $\tilde{\delta}_{1,2}=\tilde{v}_{1,2}-v\sim O(\mu) $. $\delta_i,\tilde{\delta}_{1,2}$ independent of $v$. Carrying out $\sum_v$, the first term in Eq.\Ref{derivativeC} becomes: 
\be
&&\mu s_U\sum_{\a,\b,J,K, N^\pm_J,M^\pm_K}\sum_i\frac{C_{v'-\delta_i}}{H_{v'-\delta_i}}\frac{\partial F^{\a\b}_{N^\pm_J,M^\pm_K}\lt(\overrightarrow{v'-\delta_i}\rt)}{\partial \Fe^I_a(v')}\Delta_{J,N^\pm_J}\ff_\a(v'-\delta_i+\tilde{\delta}_{1})\,\Delta_{K,M^\pm_K}\ff_\b(v'-\delta_i+\tilde{\delta}_{2})\nonumber\\
&=&\mu^3s_U\sum_{\a,\b,J,K, N^\pm_J,M^\pm_J}\frac{\cc(v')}{h(v')}\frac{\partial \cf^{\a\b}_{N^\pm_J,M^\pm_K}\lt(v'\rt)}{\partial E^I_a(v')}(N_J^++N_J^-)(M_K^++M_K^-)\partial_J f_\a(v')\partial_K f_\b(v')\nonumber\\
&&+\ O(\mu^4),\label{star1}
\ee
where $F^{\a\b}_{N^\pm_J,M^\pm_K}\lt(\overrightarrow{v'-\delta_i}\rt)$ is from the expansion of $C_{v'-\delta_i}$. Note that all vertices in $\overrightarrow{v'-\delta_i}$ are inside $U$. $F^{\a\b}_{N^\pm_J,M^\pm_K}(\vec{v})$ is a polynomial of $\ff_\a({v}_i)$. Derivatives $\partial F^{\a\b}_{N^\pm_J,M^\pm_K}/\partial \Fe^I_a$ have continuum limit ${\partial \cf^{\a\b}_{N^\pm_J,M^\pm_K}}/{\partial E^I_a}$. Thanks to summing over all $v\in U$, $\sum_i$ in Eq.\Ref{star1} sums over vertices $v'-\delta_i$ at which ${\partial F^{\a\b}_{N^\pm_J,M^\pm_K}(\overrightarrow{v'-\delta_i})}/{\partial \Fe^I_a(v')} $ are nonzero, and reduces to the Leibniz rule of ${\partial \cf^{\a\b}_{N^\pm_J,M^\pm_K}\lt(v'\rt)}/{\partial E^I_a(v')}$. 

In the second term in the result of Eq.\Ref{derivativeC}, $\delta_{v',\tilde{v}_1\pm N^\pm_J\mu\hat{J}}$ restricts $v=v'-\tilde{\delta}_1\mp N^\pm_J\mu\hat{J}\equiv v_J^\pm$. Carrying out $\sum_v$ in the second term in Eq.\Ref{derivativeC} gives 
\be
&&\mu s_U \sum_{\a,\b,J,K, N^\pm_J,M^\pm_K}\Bigg[\frac{C_{v_J^+}}{H_{v_J^+}}F^{\a\b}_{N^\pm_J,M^\pm_K}\lt(\overrightarrow{v_J^+}\rt)\Delta_{K,M^\pm_K}\ff_\b\lt(v_J^++\tilde{\delta}_2\rt)\nonumber\\
&&\  -\frac{C_{v_J^-}}{H_{v^-_J}}F^{\a\b}_{N^\pm_J,M^\pm_K}\lt(\overrightarrow{v_J^-}\rt)\Delta_{K,M^\pm_K}\ff_\b\lt(v_J^-+\tilde{\delta}_2\rt)\Bigg]\frac{\partial \ff_\a(v')}{\partial \Fe^I_a(v')}\nonumber\\
&=&-\mu^3s_U\sum_{\a,\b,J,K, N^+_J,N^-_J}(N_J^++N_J^-)(M_K^++M_K^-)\partial_J\lt[\frac{\cc}{h}\cf^{\a\b}_{N^\pm_J,M^\pm_K}\partial_K f_\b\rt](v')\frac{\partial f_\a(v')}{\partial E^I_a(v')}\nonumber\\
&&+O(\mu^4).\label{star2}
\ee

The third and fifth terms in Eq.\Ref{derivativeC} are treated similar to the second term, while the fourth and sixth terms are treated similar to the first term. As results,
\be
\text{3rd term}&=&-\mu^3s_U\sum_{\a,\b,J,K, N^+_J,N^-_J}(N_J^++N_J^-)(M_K^++M_K^-)\partial_K\lt[\frac{\cc}{h}\cf^{\a\b}_{N^\pm_J,M^\pm_K}\partial_J f_\a\rt](v')\frac{\partial f_\b(v')}{\partial E^I_a(v')}\nonumber\\
&&+O(\mu^4)\nonumber\\
\text{4th term}&=&\mu^3s_U\sum_{\a,J, N^+_J,N^-_J}\frac{\cc(v')}{h(v')}\frac{\partial \cf^\a_{N^+_J,N^-_J}\lt(v'\rt)}{\partial E^I_a(v')}(N_J^++N_J^-)\partial_J f_\a(v')+O(\mu^4)\nonumber\\
\text{5th term}&=&-\mu^3s_U\sum_{\a,J, N^+_J,N^-_J}(N_J^++N_J^-)\partial_J\lt[\frac{\cc(v')}{h(v')} \cf^\a_{N^+_J,N^-_J}\lt(v'\rt)\rt]\frac{\partial f_\a(v')}{\partial E^I_a(v')}+O(\mu^4)\nonumber\\
\text{6th term}&=&\mu^3\frac{\cc(v')}{h(v')}\frac{\partial \cf\lt(v'\rt)}{\partial E^I_a(v')}+O(\mu^4).\label{star3}
\ee

On the other hand, we apply the functional derivative to $\cc$ using Eq.\Ref{typicalcon}:
\be
&&\int_U\rmd^3\sig\frac{\cc(\sig)}{h(\sig)}\frac{\delta \cc(\sig)}{\delta E^I_a(v')}\nonumber\\
&=&\sum_{\a,\b,J,K, N^\pm_J,M^\pm_J}\frac{\cc(v')}{h(v')}\frac{\partial \cf^{\a\b}_{N^\pm_J,M^\pm_K}\lt(v'\rt)}{\partial E^I_a(v')}(N_J^++N_J^-)(M_K^++M_K^-)\partial_J f_\a(v')\partial_K f_\b(v')\nonumber\\
&&\ -\sum_{\a,\b,J,K, N^+_J,N^-_J}(N_J^++N_J^-)(M_K^++M_K^-)\partial_J\lt[\frac{\cc}{h}\cf^{\a\b}_{N^\pm_J,M^\pm_K}\partial_K f_\b\rt](v')\frac{\partial f_\a(v')}{\partial E^I_a(v')}\nonumber\\
&&\ -\sum_{\a,\b,J,K, N^+_J,N^-_J}(N_J^++N_J^-)(M_K^++M_K^-)\partial_K\lt[\frac{\cc}{h}\cf^{\a\b}_{N^\pm_J,M^\pm_K}\partial_J f_\a\rt](v')\frac{\partial f_\b(v')}{\partial E^I_a(v')}\nonumber\\
&&+\ \sum_{\a,J, N^+_J,N^-_J}\frac{\cc(v')}{h(v')}\frac{\partial \cf^\a_{N^+_J,N^-_J}\lt(v'\rt)}{\partial E^I_a(v')}(N_++N_-)\partial_J f_\a(v')\nonumber\\
&&\ -\sum_{\a,J, N^+_J,N^-_J}(N_++N_-)\partial_J\lt[\frac{\cc(v')}{h(v')} \cf^\a_{N^+_J,N^-_J}\lt(v'\rt)\rt]\frac{\partial f_\a(v')}{\partial E^I_a(v')}+ \frac{\cc(v')}{h(v')}\frac{\partial \cf\lt(v'\rt)}{\partial E^I_a(v')}.\label{fder}
\ee
Comparing Eq.\Ref{fder} with \Ref{star1} - \Ref{star3}, we obtain the following result
\be
\sum_{v\in V(\g)} s_v\frac{C_v}{H_v}\frac{\partial C_v}{\partial \Fe^I_a(v')}=\mu^3\int_U\rmd^3\sig\,s_U\frac{\cc(\sig)}{h(\sig)}\frac{\delta \cc(\sig)}{\delta E^I_a(v')}+O(\mu^4).\label{dercon}
\ee

The derivation of Eq.\Ref{dercon} only uses general patterns of $C_v$, $C_{j,v}$ in Eq.\Ref{typical} and their continuum limit, so can be easily generalized to $\sum_v \frac{C_{b,v}}{H_v}\frac{\partial C_{b,v}}{\partial \Fe^I_a(v')}$ and derivatives with respect to $\Fa^a_I(v')$. Therefore
\be
\frac{\partial{\bf H}}{\partial \Fe^I_a(v')}&=&\mu^3\int_U\rmd^3\sig\,s_U\lt[\frac{\cc(\sig)}{h(\sig)}\frac{\delta \cc(\sig)}{\delta E^I_a(v')}-\frac{\a}{4}\sum_{b=1}^3\frac{\cc_b(\sig)}{h(\sig)}\frac{\delta \cc_b(\sig)}{\delta E^I_a(v')}\rt]+O(\mu^4)\nonumber\\
&=&\mu^3\frac{\delta }{\delta E^I_a(v')}\int_\cs\rmd^3\sig \sqrt{\lt|\cc(\sig)^2-\frac{\a}{4}\sum_{b=1}^3\cc_{b}(\sig)^2\rt|}+O(\mu^4),\\
\frac{\partial{\bf H}}{\partial \Fa_I^a(v')}&=&\mu^3\int_U\rmd^3\sig\, s_U\lt[\frac{\cc(\sig)}{h(\sig)}\frac{\delta \cc(\sig)}{\delta A_I^a(v')}-\frac{\a}{4}\sum_{b=1}^3\frac{\cc_b(\sig)}{h(\sig)}\frac{\delta \cc_b(\sig)}{\delta A_I^a(v')}\rt]+O(\mu^4),\\
&=&\mu^3\frac{\delta }{\delta A_I^a(v')}\int_\cs\rmd^3\sig \sqrt{\lt|\cc(\sig)^2-\frac{\a}{4}\sum_{b=1}^3\cc_{b}(\sig)^2\rt|}+O(\mu^4).
\ee
$\int_U$ can be replaced by $\int_\cs$ because the functional derivative is local. This result shows that the lattice continuum limit of partial derivatives in discrete variables gives the functional derivatives in smooth fields. 

Recall Eqs.\Ref{kincon} and \Ref{dercon02}, we obtain the lattice continuum limit of discrete semiclassical EOMs \Ref{eom0}:
\be
-\frac{\rmd A^a_I(v)}{\rmd \t}&=&\frac{\kappa\b}{2}\frac{\delta }{\delta E^I_a(v)}\int_\cs\rmd^3\sig \sqrt{\lt|\cc(\sig)^2-\frac{\a}{4}\sum_{a=1}^3\cc_{a}(\sig)^2\rt|}+O(\mu),\label{clim1}\\
\frac{\rmd E^I_a(v)}{\rmd \t}&=&\frac{\kappa\b}{2}\frac{\delta }{\delta A_I^a(v)}\int_\cs\rmd^3\sig \sqrt{\lt|\cc(\sig)^2-\frac{\a}{4}\sum_{a=1}^3\cc_{a}(\sig)^2\rt|}+O(\mu).\label{clim2}
\ee
The result recovers the classical EOMs \Ref{hamitoncon0} of the gravity-dust system in the continuum when $\cc(\sig)^2-\frac{\a}{4}\sum_{a=1}^3\cc_{a}(\sig)^2>0$.

The above derivation replies on the assumption that $v'\in U$, $r(v',\partial U)\gg\mu$, and $s_v=s_U$ is constant on $U$. But if we violate this assumption, i.e. let $v'\in U$, $r(v',\partial U)\sim\mu$, and $s_v$ changes sign outside $U$, then in the lattice continuum limit $\mu\to0$, $v'$ belongs to the boundary where $s_v$ jumps and $\cc(\sig)^2-\frac{\a}{4}\sum_{a=1}^3\cc_{a}(\sig)^2=0$. Semiclassical EOMs at this $v'$ cannot relate to Eqs.\Ref{clim1} and \Ref{clim2} by the lattice continuum limit, because the functional derivative is ill-defined at $v'$.

In our quantization, nonholonomic constraints $\cc(\sig)^2-\frac{\a}{4}\sum_{a=1}^3\cc_{a}(\sig)^2>0$ and $\cc<0$ are not imposed to the Hilbert space $\ch_\g$. Therefore ${\bf H}$ are defined on the entire phase space $\cp_\g$, thus the continuum limit Eqs.\Ref{clim1} and \Ref{clim2} extend the continuum theory to the regime where nonholonomic constraints are not valid. The relation between Eqs.\Ref{clim1} - \Ref{clim2} and the classical EOMs \Ref{hamitoncon0} is sensitive to the choice of initial condition. Here the initial condition is given by $[g']$ at which the initial coherent state $\Psi^t_{[g']}$ is peaked.  $\Psi^t_{[g']}$ is semiclassical if $[g']$ is in the classical allowed regime of the phase space, while the classical allowed regime satisfies the non-holonomic constraints required by the classical gravity-dust system. Eqs.\Ref{clim1} and \Ref{clim2} indeed coincide with classical EOMs \Ref{hamitoncon0} of the continuum theory, if the initial data $g'$ satisfies (discretized) nonholonomic constraints: 

\begin{itemize}

\item For gravity coupled to Brown-Kucha\v{r} dust, if the initial data $g'$ at $\t=0$ satisfies $C_v^2-\frac{1}{4}\sum_{a=1}^3C_{a,v}^2>0$ and ${C_v}<0$ at all $v\in V(\g)$, these two non-holonomic constraints are going to be still satisfied by the solution to EOMs \Ref{clim1} and \Ref{clim2} within a finite time period $\t\in[0,T_0]$, simply because the solution is a continuous function in $\t$. Therefore $|\ |$ in \Ref{clim1} and \Ref{clim2} can be removed at least with in this time period. 

On the other hand, although $C_v^2-\frac{1}{4}\sum_{a=1}^3C_{a,v}^2$ is not exactly conserved in \Ref{eom0} (or \Ref{hamiton}) due to the anomaly from discretization \cite{Giesel:2007wn}, it is approximately conserved up to $O(\mu)$ because its continuum limit $\cc^2-\frac{1}{4}\sum_{a=1}^3\cc_{a}^2$ is conserved by the continuum limit Eqs.\Ref{clim1} and \Ref{clim2}. ${C_v}$ cannot flip sign by the similar reason. Therefore $C_v^2-\frac{1}{4}\sum_{a=1}^3C_{a,v}^2>0$ and ${C_v}<0$ can continuously be satisfied by the solution at and even after $T_0$. By adding another time period $[T_0,2T_0]$, repeating the argument iteratively, we can extend the time period to entire $[0,T]$ in which $C_v^2-\frac{1}{4}\sum_{a=1}^3C_{a,v}^2>0$ and ${C_v}<0$ are satisfied, when $\mu$ is sufficiently small\footnote{$T\to\infty$ is more subtle because accumulating errors of $O(\mu)$ over infinite amount of time might cause a finite change of $C_v^2-\frac{1}{4}\sum_{a=1}^3C_{a,v}^2$ and flip the sign.}.
Then semiclassical EOMs from $A_{[g],[g']}$ reproduce classical EOMs \Ref{hamitoncon0} for gravity coupled to Brown-Kucha\v{r} dust in the continuum limit:
\be
-\frac{\rmd A^a_I(v)}{\rmd \t}&=&\frac{\kappa\b}{2}\frac{\delta }{\delta E^I_a(v)}\int_\cs\rmd^3\sig \sqrt{\cc(\sig)^2-\frac{1}{4}\sum_{a=1}^3\cc_{a}(\sig)^2}+O(\mu),\label{clim12}\\
\frac{\rmd E^I_a(v)}{\rmd \t}&=&\frac{\kappa\b}{2}\frac{\delta }{\delta A_I^a(v')}\int_\cs\rmd^3\sig \sqrt{\cc(\sig)^2-\frac{1}{4}\sum_{a=1}^3\cc_{a}(\sig)^2}+O(\mu).\label{clim22}
\ee

\item A similar reasoning applies to gravity coupled to Gaussian dust, when the initial data $g'$ of $A_{[g],[g']}$ satisfy $C_v<0$ and $C_{a,v}=0$, both $C_v$ and $C_{a,v}$ are approximately conserved if $\mu$ is sufficiently small, since they are conserved in the continuum limit, thus $C_v<0$ is preserved by the time evolution for sufficiently small $\mu$. Then semiclassical EOMs of reduced phase space LQG with Gaussian dust reproduce classical EOMs \Ref{hamitoncon0} in the continuum limit up to a flip of time direction:
\be
\frac{\rmd A^a_I(v)}{\rmd \t}&=&\frac{\kappa\b}{2}\frac{\delta }{\delta E^I_a(v)}\int_\cs\rmd^3\sig\ \cc(\sig)+O(\mu),\label{clim13}\\
-\frac{\rmd E^I_a(v)}{\rmd \t}&=&\frac{\kappa\b}{2}\frac{\delta }{\delta A_I^a(v')}\int_\cs\rmd^3\sig\ \cc(\sig)+O(\mu).\label{clim23}
\ee
Recall that time direction has been flipped to flow backward in Section \ref{BK} in order to obtain a positive Hamiltonian.

\item If the initial data does not satisfy nonholonomic constraints, $\Psi^t_{[g']}$ is not anymore semiclassical. The continuum limit of semiclassical EOMs derived from $A_{[g],[g']}$ cannot be related to classical EOMs \Ref{hamitoncon0} of the gravity-dust system. Existence of non-classical solutions has been anticipated in \cite{Giesel:2007wn}, and viewed as analogs of negative energy states in relativistic QFT, because when Eq.\Ref{P=-h} is viewed as constraint, it can be written as ${P}^2+({\cc}-{q}^{\a\b}{\cc}_\a{\cc}_\b)=0$ whose quantization would be an analog of Klein-Gordan operator. But that non-classical solutions appear or disappear is determined by initial conditions, similar to the situation of negative energy states in QFT.

\end{itemize}

Some examples of solutions of semiclassical EOMs and their continuum limit are studied in cosmological perturbation theory in \cite{cospert}.

\section{Asymptotics of Transition Amplitude}

Assuming initial and final states $\Psi_{[g']}^t,\Psi_{[g]}^t$ are both semiclassical in the sense that both $[g'],[g]$ are within the classical allowed regime, if $[g],[g']$ are connected by the trajectory $g(\t)$ satisfying Eqs.\Ref{hamiton4.11}, as $t\to0$, integrals $\int\prod_{i=1}^{N+1}\rmd g_i$ in the path integral \Ref{Agg} dominate at this semiclassical trajectory:
\be
\frac{A_{[g],[g']}}{\|\Psi^t_{[g]}\|\,\|\Psi^t_{[g']}\|}=\int\rmd h\, \frac{(2\pi t)^{\cn/2}}{\sqrt{\det(-H)}}\,\nu[g(\t),h] \, e^{S[g(\t),h]/t}\,\lt[1+O(t)\rt],\label{asymp1}
\ee 
where $\cn$ is the total dimension of the integral $\int\prod_{i=1}^{N+1}\rmd g_i$ in Eq.\Ref{Agg}, and $H$ is the Hessian matrix at the solution. $g(\t)$ is unique up to SU(2) gauge transformations. $S[g(\t),h]$ is the action evaluated at the solution, where the continuous trajectory $g(\t)\simeq g_i$ approximates the discrete solution as $\Delta\t$ small. Here we still have $\int \rmd h=\int\prod_v\rmd\mu_H( h_v)$ which integrates gauge transformations $g'\to g'^h$ of the initial data. 

If the initial and final data $[g'],[g]$ are not connected by any trajectory $g(\t)$ satisfying Eqs.\Ref{hamiton4.11}, the amplitude is suppressed as $t\to0$:
\be
\frac{A_{[g],[g']}}{\|\Psi^t_{[g]}\|\,\|\Psi^t_{[g']}\|}=O(t^M),\quad \forall\ M>0.
\ee

\section{Comparison with Spin Foam Formulation and Outlook}\label{Comparison with Spin Foam Formulation}

The above analysis demonstrates the semiclassical consistency of the new path integral formulation from reduced phase space LQG. If we compare our results to the spin foam formulation, we find following advantages of our path integral formulation:

\begin{enumerate}

\item Our path integral formulation is free of the cosine problem. The initial condition $[g']$ given by the semiclassical initial state $\Psi_{[g']}^t$ determines a unique solution of semiclassical EOMs up to SU(2) gauge freedom. Therefore the asymptotic formula \Ref{asymp1} has only a single exponential in the integrand.

A key reason why we obtain unique solution and avoid the cosine problem is that all solutions of discrete EOMs \Ref{eoms1} and \Ref{eoms2} admit the time continuous limit. If spin foam formulation admitted the time continuous limit or anything similar, the continuous time EOMs (critical equations) would have forbidden the 4d orientation to jump, and suppressed contributions from orientation-changing evolutions to spin foam amplitude. 

\item Our path integral formulation is free of the flatness problem. The semiclassical analysis of the path integral has been shown to reproduce the classical EOMs \Ref{hamitoncon0}, which are Einstein equation formulated in the reduced phase space. Semiclassical EOMs \Ref{hamiton} admit all curved solutions that are physically interesting. For instance, \cite{Han:2019vpw} has demonstrated the homogeneous and isotropic cosmology as a solution, while \cite{cospert} obtains cosmological perturbation theory from solutions. Note that the flat spacetime is not a solution of semiclassical EOMs because of the presence of physical dust field with positive energy density.

\item There is a clear link between our path integral formulation and the canonical LQG. The path integral \Ref{Agg} is rigorously derived from the canonical formulation in the reduced phase space. The unitarity is manifest because the path integral is the transition amplitude of unitary evolution generated by the Hamiltonian $\hat{\bf H}$.

\item The path integral formla \Ref{Agg} is clearly finite (irrelevant to the cosmological constant), because of the transition amplitude $A_{[g],[g']}=\langle \Psi^t_{[g]}|\,\exp[-\frac{i}{\hbar}T \hat{\bf H}]\,|\Psi^t_{[g']}\rangle$ is finite. All ingredients $\Psi^t_{[g]},\ \Psi^t_{[g']},\ \exp[-\frac{i}{\hbar}T \hat{\bf H}]$, and $\langle\ \cdot\ |\ \cdot\ \rangle$ are well defined. 

\end{enumerate}

Our formulation may still have issues of computational complexity and lattice dependence similar to the spin foam formulation, at least at the present stage. However studies of the new path integral formulation are still at very preliminary stage, and research on overcoming these issues will be carried out in the future.

\begin{enumerate}

\item At the level of discrete path integral \Ref{Agg}, the action $S[g,h]$ depends on the non-polynomial operator $\hat{\bf H}$ and its matrix element, which is hard to compute. However because $\Delta\t$ is arbitrarily small, we may consider a formal time continuous limit at the level of path integral, as in the standard QFT. The resulting path integral formula integrates over continuous paths, then the matrix element of $\hat{\bf H}$ in $S[g,h]$ reduces to the coherent state expectation value $\langle \psi_g^t | \hat{\bf H}|\psi_g^t\rangle$, which is computable as a perturbative expansion in $t$ by using the method in \cite{Giesel:2006um}. Therefore perturbative techniques in QFT (more precisely, the lattice perturbation theory) should be applied to our path integral formulation to compute quantities such as correlation functions and quantum effective action as power expansions in $t$.  

\item Our path integral formulation depends on the cubic lattice $\g$ even after taking the time continuous limit. Currently the lattice continuum limit at the quantum level is not clear for our formulation (in Section \ref{Lattice Continuum Limit}, the lattice continuum limit $\mu\to0$ is taken after the semiclassical limit $\mu\to0$). We expect to see effects of lattice continuum limit order by order in $t$ in perturbative computations. 

\end{enumerate}


\section*{Acknowledgements}

This work receives support from the National Science Foundation through grant PHY-1912278. Mathematica computations in this work are carried out on the HPC server at Fudan University and the KoKo HPC server at Florida Atlantic University. The authors acknowledge Ling-Yan Hung for sharing the computational resource at Fudan University. 
	

\appendix

\section{Proof of Eq.\Ref{typical}}\label{typeexplanation}

There are 2 useful properties of $C_v$ and $C_{a,v}$

\begin{itemize}

\item $C_v,\ C_{a,v}$ are polynomials of $h(e),\ p^a(e)$ and $Q_v^{1/2}$. By applying Eqs.\Ref{interffff1}, \Ref{interffff1} and expand in $\mu$, $C_v,\ C_{a,v}$ become series of $\mu$ and $\ff_\a(v)$. 

\item In the continuum limit $C_v=\mu^3\cc(v)+O(\mu^4)$, $C_{a,v}=\mu^3\cc_{a}(v)+O(\mu^4)$ where the leading order is of $O(\mu^3)$ and both $\cc$ and $\cc_a$ are polynomials of $f_\a$ and their 1st order derivatives\footnote{$F_{IJ}^a$ has only 1st order derivatives of $A_I^a$. $\b K^a_I=A^a_I-\G^a_I$ where $\G^a_I=\frac{1}{2} \epsilon^{a b c} E_{c}^{J}\left[E_{I, J}^{b}-E_{J, I}^{b}+E_{b}^{K} E_{I}^{d} E_{K, J}^{d}\right]+\frac{1}{4} \epsilon^{a b c} E_{c}^{J}\left[2 E_{I}^{b} \frac{(\operatorname{det}(E))_{, J}}{\operatorname{det}(E)}-E_{J}^{b} \frac{(\operatorname{det}(E))_{, I}}{\operatorname{det}(E)}\right]$. Here $\det(E(v))=q(v)$, and the inverse $E^a_I$ is a polynomial of $E^I_a$ and $\det(E)^{-1}$. }. Each term in $\cc$ and $\cc_a$ contain no more than 2 derivatives.

\end{itemize}



We extract arbitrarily two terms at $O(\mu^n)$ in the expansion of $C_v$ and $C_{a,v}$. Generically they may be written as 
\be
&&\mathfrak{f}_{1}\left(v_{1}\right)\mathrm{\mathrm{\mathfrak{f}}}_{2}\left(v_{2}\right)\ldots\mathrm{\mathfrak{f}}_{n}\left(v_{n}\right)\mathrm{\mathfrak{f}}_{n+1}\left(v_{n+1}\right)\cdots\mathrm{\mathfrak{f}}_{m}\left(v_{m}\right)\nonumber\\
\text{and}&& \mathrm{\mathrm{\mathfrak{f}}}_{1}\left(v_{1}^{\prime}\right)\mathrm{\mathrm{\mathfrak{f}}}_{2}\left(v_{2}^{\prime}\right)\ldots\mathrm{\mathfrak{f}}_{n}\left(v_{n}^{\prime}\right)\mathrm{\mathfrak{f'}}_{n+1}\left(v'_{n+1}\right)\cdots\mathrm{\mathfrak{f'}}_{q}\left(v'_{q}\right).
\ee
They may share $\ff_1,\cdots,\ff_n$ although locations of $\ff_1,\cdots,\ff_n$, $v_i$ and $v_i'$, may be different between these 2 terms. Distances from $v$ to $v_i,v_i'$ are of $O(\mu)$. $\ff_i$ and $\ff_i'$ are factors not shared by these 2 terms. If the relative sign between these 2 terms is negative, we can perform the following reduction
\be	
&&\mathfrak{f}_{1}\left(v_{1}\right)\mathrm{\mathrm{\mathfrak{f}}}_{2}\left(v_{2}\right)\ldots\mathrm{\mathfrak{f}}_{n}\left(v_{n}\right)\mathrm{\mathfrak{f}}_{n+1}\left(v_{n+1}\right)\cdots\mathrm{\mathfrak{f}}_{m}\left(v_{m}\right)-\mathrm{\mathrm{\mathfrak{f}}}_{1}\left(v_{1}^{\prime}\right)\mathrm{\mathrm{\mathfrak{f}}}_{2}\left(v_{2}^{\prime}\right)\ldots\mathrm{\mathfrak{f}}_{n}\left(v_{n}^{\prime}\right)\mathrm{\mathfrak{f'}}_{n+1}\left(v'_{n+1}\right)\cdots\mathrm{\mathfrak{f'}}_{q}\left(v'_{q}\right)\nonumber
\\
&=&\mathfrak{f}_{1}\left(v_{1}\right)\mathrm{\mathrm{\mathfrak{f}}}_{2}\left(v_{2}\right)\ldots\mathrm{\mathfrak{f}}_{n}\left(v_{n}\right)\mathrm{\mathfrak{f}}_{n+1}\left(v_{n+1}\right)\cdots\mathrm{\mathfrak{f}}_{m}\left(v_{m}\right)-\mathrm{\mathrm{\mathfrak{f}}}_{1}\left(v_{1}^{\prime}\right)\mathrm{\mathrm{\mathfrak{f}}}_{2}\left(v_{2}^{\prime}\right)\ldots\mathrm{\mathfrak{f}}_{n}\left(v_{n}^{\prime}\right)\mathrm{\mathfrak{f}}_{n+1}\left(v_{n+1}\right)\cdots\mathrm{\mathfrak{f}}_{m}\left(v_{m}\right)\nonumber\\
&&+\ \mathrm{\mathrm{\mathfrak{f}}}_{1}\left(v_{1}^{\prime}\right)\mathrm{\mathrm{\mathfrak{f}}}_{2}\left(v_{2}^{\prime}\right)\ldots\mathrm{\mathfrak{f}}_{n}\left(v_{n}^{\prime}\right)\mathrm{\mathfrak{f}}_{n+1}\left(v_{n+1}\right)\cdots\mathrm{\mathfrak{f}}_{m}\left(v_{m}\right)-\mathrm{\mathrm{\mathfrak{f}}}_{1}\left(v_{1}^{\prime}\right)\mathrm{\mathrm{\mathfrak{f}}}_{2}\left(v_{2}^{\prime}\right)\ldots\mathrm{\mathfrak{f}}_{n}\left(v_{n}^{\prime}\right)\mathrm{\mathfrak{f'}}_{n+1}\left(v'_{n+1}\right)\cdots\mathrm{\mathfrak{f'}}_{q}\left(v'_{q}\right)\nonumber\\
&=&\lt[\mathfrak{f}_{1}\left(v_{1}\right)\mathrm{\mathrm{\mathfrak{f}}}_{2}\left(v_{2}\right)\ldots\mathrm{\mathfrak{f}}_{n}\left(v_{n}\right)-\mathrm{\mathrm{\mathfrak{f}}}_{1}\left(v_{1}^{\prime}\right)\mathrm{\mathrm{\mathfrak{f}}}_{2}\left(v_{2}^{\prime}\right)\ldots\mathrm{\mathfrak{f}}_{n}\left(v_{n}^{\prime}\right)\rt]\mathrm{\mathfrak{f}}_{n+1}\left(v_{n+1}\right)\cdots\mathrm{\mathfrak{f}}_{m}\left(v_{m}\right)\nonumber\\
&&+\ \mathrm{\mathrm{\mathfrak{f}}}_{1}\left(v_{1}^{\prime}\right)\mathrm{\mathrm{\mathfrak{f}}}_{2}\left(v_{2}^{\prime}\right)\ldots\mathrm{\mathfrak{f}}_{n}\left(v_{n}^{\prime}\right)\left[\mathrm{\mathfrak{f}}_{n+1}\left(v_{n+1}\right)\cdots\mathrm{\mathfrak{f}}_{m}\left(v_{m}\right)-\mathrm{\mathfrak{f'}}_{n+1}\left(v'_{n+1}\right)\cdots\mathrm{\mathfrak{f'}}_{q}\left(v'_{q}\right)\right],\label{2squarebrackets}
\ee
The quantity in the 1st square bracket of the above result is the difference of two monomials $\ff_{1}(v_{1})\ff_{2}(v_{2})\dots \ff_{n}(v_{n})$ and $\ff_{1}(v'_{1})\ff_{2}(v'_{2})\dots \ff_{n}(v'_{n})$ sharing the same set of $\ff_{1,\cdots,n}$, and can be further reduced
\be
&&\ff_{1}(v_{1})\ff_{2}(v_{2})\dots \ff_{n}(v_{n})-\ff_{1}(v'_{1})\ff_{2}(v'_{2})\dots \ff_{n}(v'_{n})	\nonumber\\
&=&\ff_{1}(v_{1})\ff_{2}(v_{2})\dots \ff_{n}(v_{n})-\ff_{1}(v'_{1})\ff_{2}(v'_{2})\dots \ff_{n}(v'_{n})+\ff_{1}(v{}_{1})\ff_{2}(v'_{2})\dots \ff_{n}(v'_{n})-\ff_{1}(v{}_{1})\ff_{2}(v'_{2})\dots \ff_{n}(v'_{n})\nonumber\\
&=&\ff_{1}(v_{1})\left[\ff_{2}(v_{2})\dots \ff_{n}(v_{n})-\ff_{2}(v'_{2})\dots \ff_{n}(v'_{n})\right]+\left[\ff_{1}(v_{1})-\ff_{1}(v'_{1})\right]\ff_{2}(v'_{2})\dots \ff_{n}(v'_{n})\nonumber\\
&=&\cdots\nonumber\\
&=&\sum_{i=1}^n\ff_{1}(v_{1})\cdots\ff_{i-1}(v_{i-1})\left[\ff_{i}(v_{i})-\ff_{i}(v'_{i})\right]\ff_{i+1}(v'_{i+1})\dots \ff_{n}(v'_{n}).
\ee
Inserting this result back into Eq.\Ref{2squarebrackets} gives
\be
&&\mathfrak{f}_{1}\left(v_{1}\right)\mathrm{\mathrm{\mathfrak{f}}}_{2}\left(v_{2}\right)\ldots\mathrm{\mathfrak{f}}_{n}\left(v_{n}\right)\mathrm{\mathfrak{f}}_{n+1}\left(v_{n+1}\right)\cdots\mathrm{\mathfrak{f}}_{m}\left(v_{m}\right)-\mathrm{\mathrm{\mathfrak{f}}}_{1}\left(v_{1}^{\prime}\right)\mathrm{\mathrm{\mathfrak{f}}}_{2}\left(v_{2}^{\prime}\right)\ldots\mathrm{\mathfrak{f}}_{n}\left(v_{n}^{\prime}\right)\mathrm{\mathfrak{f'}}_{n+1}\left(v'_{n+1}\right)\cdots\mathrm{\mathfrak{f'}}_{q}\left(v'_{q}\right)\nonumber\\
&=&\sum_{i=1}^n\ff_{1}(v_{1})\cdots\ff_{i-1}(v_{i-1})\left[\ff_{i}(v_{i})-\ff_{i}(v'_{i})\right]\ff_{i+1}(v'_{i+1})\dots \ff_{n}(v'_{n})\nonumber\\
&&+\ \mathrm{\mathrm{\mathfrak{f}}}_{1}\left(v_{1}^{\prime}\right)\mathrm{\mathrm{\mathfrak{f}}}_{2}\left(v_{2}^{\prime}\right)\ldots\mathrm{\mathfrak{f}}_{n}\left(v_{n}^{\prime}\right)\left[\mathrm{\mathfrak{f}}_{n+1}\left(v_{n+1}\right)\cdots\mathrm{\mathfrak{f}}_{m}\left(v_{m}\right)-\mathrm{\mathfrak{f'}}_{n+1}\left(v'_{n+1}\right)\cdots\mathrm{\mathfrak{f'}}_{q}\left(v'_{q}\right)\right],
\ee
while there is no reduction for the 2nd square bracket. Here the point of this reduction is to manifest the difference $\ff_{i}(v_{i})-\ff_{i}(v'_{i})$ in the formula.

We insert the above result back into $C_v$ and $C_{v,a}$ so that they become polynomials of $\ff_\a$ and $\Delta\ff_\a(v,v')\equiv\ff_{\a}(v)-\ff_{\a}(v')$. We make further similar reduction as above, by including $\Delta\ff_{\a}$ as one of generators of the polynomial. As a result from iteration, we obtain at $O(\mu^n)$
\be
&&\mu^n\lt[\mathrm{Pol}_n(\ff_\a )+\sum_{p>0}\mathrm{Pol}^p_n(\ff_\a,\Delta\ff_\a )+\sum_{k\geq0,l>0}\mathrm{Pol}^{k,l}_n(\ff_\a,\Delta\ff_\a,\Delta^2\ff_\a )\rt]\nonumber\\
&=&\mu^n\lt[\mathrm{Pol}_n(\ff_\a )+\sum_{p>0}\mu^p\mathrm{Pol}_n^p\lt(\ff_\a,\Delta\ff_\a/\mu \rt)+\sum_{k\geq0,l>0}\mu^{k+2l}\mathrm{Pol}_n^{k,l}\lt(\ff_\a,\Delta\ff_\a/\mu,\Delta^2\ff_\a/\mu^2 \rt)\rt]
\ee
$\Delta^2\ff_\a=\Delta\ff_\a(v,v')-\Delta\ff_\a(\tilde{v},\tilde{v}')$. $\Delta\ff_\a/\mu,\Delta^2\ff_\a /\mu^2$ are lattice derivatives. $\mathrm{Pol}_n(\ff_\a )$ is a polynomial of $\ff_\a$. $\mathrm{Pol}_n^p(\ff_\a,\Delta\ff_\a )$ is a polynomial homogeneous in $\Delta\ff_\a$ of degree $p$. $\mathrm{Pol}_n^{k,l}(\ff_\a,\Delta\ff_\a,\Delta^2\ff_\a )$ is a polynomial homogeneous in $\Delta\ff_\a$ and $\Delta^2\ff_\a$ of degree $k$ and $l$ respectively.  

When $\mathrm{Pol}_n(\ff_\a )$,  $\mathrm{Pol}_n^p(\ff_\a,\Delta\ff_\a )$, and $\mathrm{Pol}_n^{k,l}(\ff_\a,\Delta\ff_\a,\Delta^2\ff_\a )$ are nonzero, their continuum limits do not vanish, because otherwise they can be further reduced to higher order in $\Delta\ff_\a$.

We are interested in expansions of $C_v$ and $C_{v,a}$ truncated up to $O(\mu^3)$ to be relevant to their continuum limit. So we consider
\be
n\leq 3,\quad n+p\leq 3,\quad n+k+2l\leq 3.
\ee
Continuum limits of $C_v$ and $C_{v,a}$ contain no term of 3 derivatives, so
\be
k=0,\quad p\leq 2.
\ee 
Moreover $C_v,C_{v,a}\sim \mu^3$ in the continuum limit. So at $n=0$, $\mathrm{Pol}_0(\ff_\a ),\ \mathrm{Pol}_0^p(\ff_\a,\Delta\ff_\a )$, and $\mathrm{Pol}_0^{k,l}(\ff_\a,\Delta\ff_\a,\Delta^2\ff_\a )$ have to vanish, since otherwise they can produce nonzero continuum limit at $O(\mu^0),\ O(\mu^1),\ O(\mu^2)$ 
\be
\mathrm{Pol}_0(f_\a )+\sum_{p=1}^2\mu^p\mathrm{Pol}_0^p\lt(f_\a,\partial f_\a \rt)+\mu^{2}\mathrm{Pol}_0^{0,1}\lt(f_\a,\partial f_\a,\partial^2\ff_\a/\mu^2 \rt)
\ee
By similar arguments, $\mathrm{Pol}_1(\ff_\a )$ and $\mathrm{Pol}_1^1(\ff_\a,\Delta\ff_\a )$ has to vanish at $n=1$, and $\mathrm{Pol}_2(\ff_\a )$ has to vanish at $n=2$. As a result, $C_v$ and $C_{v,a}$ can be written as
\be
&&\mu\lt[\mathrm{Pol}_1^2\lt(\ff_\a,\Delta\ff_\a \rt)+\mathrm{Pol}_1^{0,1}\lt(\ff_\a,\Delta\ff_\a,\Delta^2\ff_\a \rt)\rt]+\mu^2\mathrm{Pol}_2^1\lt(\ff_\a,\Delta\ff_\a \rt)+\mu^3\mathrm{Pol}_3(\ff_\a ) +O(\mu^4)\nonumber\\
&\to&\mu^3\lt[\mathrm{Pol}_1^2\lt(f_\a,\partial f_\a\rt)+\mathrm{Pol}_1^{0,1}\lt(f_\a,\partial f_\a,\partial^2 f_\a \rt)+\mathrm{Pol}_2^1\lt( f_\a,\partial f_\a \rt)+\mathrm{Pol}_3(f_\a )\rt] +O(\mu^4)
\ee
Recall that continuum limits of $C_v$ and $C_{v,a}$, $\cc$ and $\cc_a$, contain no second order derivative. So $\mathrm{Pol}_1^{0,1}\lt(\ff_\a,\Delta\ff_\a,\Delta^2\ff_\a \rt)$ has to vanish. Finally we obtain 
\be
C_v\ \text{or}\ C_{v,a}=\mu\mathrm{Pol}_1^2\lt(\ff_\a,\Delta\ff_\a \rt)+\mu^2\mathrm{Pol}_2^1\lt(\ff_\a,\Delta\ff_\a \rt)+\mu^3\mathrm{Pol}_3(\ff_\a ) +O(\mu^4).\label{PolsCC}
\ee

Given any $v_1,v_2$ of $O(\mu)$-distance from $v$, 
\be
v_1=v+M_1\mu\hat{1}+N_1\mu\hat{2}+P_1\mu\hat{3},\quad v_2=v+M_2\mu\hat{1}+N_2\mu\hat{2}+P_2\mu\hat{3}.
\ee
we define 
\be
v_3=v+M_1\mu\hat{1}+N_1\mu\hat{2}+P_2\mu\hat{3},\quad v_4=v+M_1\mu\hat{1}+N_2\mu\hat{2}+P_2\mu\hat{3}.
\ee
so that
\be
v_1-v_2&=&(v_1-v_3)+(v_3-v_4)+(v_4-v_2),\\
\Delta \ff_\a(v_1,v_2)&=&\Delta_3\ff_\a(v_1,v_3)+\Delta_2\ff_\a(v_3,v_4)+\Delta_1\ff_\a(v_4,v_2),\label{directeddiff}
\ee
where $\Delta_3\ff_\a(v_1,v_3),\Delta_2\ff_\a(v_3,v_4),\Delta_1\ff_\a(v_4,v_2)$ are differences along $3,2,1$ directions respectively. Inserting Eq.\Ref{directeddiff} and expand, Eq.\Ref{PolsCC} can be rewritten as
\be
C_v\ \text{or}\ C_{v,a}=\mu\mathrm{Pol}'{}_1^2\lt(\ff_\a,\Delta_J\ff_\a \rt)+\mu^2\mathrm{Pol}'{}_2^1\lt(\ff_\a,\Delta_J\ff_\a \rt)+\mu^3\mathrm{Pol}_3(\ff_\a ) +O(\mu^4)
\ee
where every difference is along 1, 2, or 3 direction.

\bibliographystyle{jhep}

\bibliography{muxin}

\end{document}